\documentclass[pra,twocolumn,amsmath,amssymb,amsfonts,longbibliography]{revtex4-1}
\usepackage{graphicx}
\usepackage{dcolumn}
\usepackage{bm}
\usepackage[colorlinks,urlcolor=blue,citecolor=blue,linkcolor=black]%
{hyperref}
\usepackage{amsmath}
\usepackage{amsfonts}
\usepackage{amssymb}
\usepackage{feynmp}
\usepackage{tabularx}
\usepackage{extarrows}%
\begin{document}
\title{Consistent bosonization-debosonization I:\\\emph{A resolution of the nonequilibrium transport puzzle}}
\author{Nayana Shah}
\affiliation{Department of Physics, University of Cincinnati, Ohio 45221-0011, USA}
\author{C.~J.~Bolech}
\affiliation{Department of Physics, University of Cincinnati, Ohio 45221-0011, USA}
\date{August 12$^\mathrm{th}$, 2015}

\begin{abstract}
We critically reexamine the bosonization-debosonization procedure for systems
including certain types of localized features (although more general scenarios
are possible). By focusing on the case of a tunneling junction out of
equilibrium, we show that the conventional approach gives results that are not
consistent with the exact solution of the problem even at the qualitative
level. We identify inconsistencies that can adversely affect the results of
all types of calculations. We subsequently show a way to avoid these and
proceed consistently. The extended framework that we develop here should be 
widely applicable.

\end{abstract}
\maketitle

\section{Introduction}

It is well known that quantum physics gets richer and more peculiar as one
considers reduced-dimensionality scenarios. Such scenarios are nowadays far
from being esoteric. From highly anisotropic and artificially layered
materials, to nanostructures, to confined ultra-cold atomic gases, examples
abound of what was once a playground for theorists but modern experimental
techniques turn into a practical reality. The physics of one-dimensional
systems provides an example in which strong quantum effects together with
interactions and restricted kinematics modify the expectations we bring with
us from our more familiar three-dimensional world \cite{Giamarchi,Gogolin}. A
case in point, the successful paradigm of the Landau Fermi liquid generically
breaks down and gives rise to a new type of quantum fluid known as the
Luttinger liquid \cite{haldane1981}. A technical stepping stone on which the
generality of this new paradigm rests is the technique known as bosonization.

The term bosonization refers to the practical possibility of describing the
excitations of fermionic systems via a description based on bosonic degrees of
freedom. The key observation is that for a fermionic one-dimensional system
with strictly linear dispersion and no cutoff, the excitations at constant
fermion number are particle-hole pairs that can be used to construct bosonic
operators which completely capture the full excitation spectrum; such a view
is known as the constructive approach \cite{vondelft1998a}. The conceptual
advantage of the constructive point of view is that it highlights the fact
that bosonization is an exact correspondence between the two systems. There
are also various complementary presentations based on the matching of
correlators and known as the field-theoretic or the hydrodynamic approaches
(these are, for example, more amenable to the conceptual description of the
phenomenology of Luttinger liquids) \cite{Giamarchi,Gogolin,shankar1995}.

Since the conditions of linear dispersion with large bandwidth and conserved
particle number are all natural approximations for systems at sufficiently low
temperatures (much lower than the bandwidth and the energy range of deviation
from linearity), the applicability of bosonization is ubiquitous for all types
of one-dimensional systems. Moreover, since within its applicability
conditions the bosonization mapping is exact, it can equally well be used in
both equilibrium and out--of-equilibrium situations. In particular, it
provides a fertile ground for the study of transport phenomena in a variety of settings.

\subsection{Transport Problems and Bosonization}

Due to the versatility of the technique, a sizable fraction of the current
theoretical studies of transport in one-dimensional settings rests on the use
of bosonization. If the applied voltages are sufficiently low (in the same
sense as discussed above for the temperature), one can use bosonization even
if some biases are large compared with other characteristic energy scales in
the system and the problem falls outside the linear-response regime.
Bosonization thus provides us in many a case with a powerful way of addressing
strongly nonequilibrium transport problems.

Given the vast array of possible experimental situations, there are many types
of setups to be considered and one has to proceed in a class by class basis
\footnote{For example, an important class of problems (which we will not study
here) is when one has a one-dimensional conductor like a nanowire attached to
electrodes with arbitrary energy distributions \cite{gutman2010a,*gutman2010b},
and as a result non-equilibrium distribution functions are imprinted into
the Luttinger liquid via the tunneling process from the contacts.
We shall neither consider at this point the case when multiple electrodes are Luttinger
liquids \cite{chamon2003,*agarwal2009,*mintchev2013,*mardanya2015}, 
to which our considerations should nevertheless be generalizable.}. 
We shall restrict ourselves to the situation in which the leads are Fermi liquids in
different equilibria (as is typical of Landauer-style setups; cf.~Refs.~%
\onlinecite{imry1999,blanter2000}%
) and the nonequilibrium situation is confined to a zero-dimensional system
(\textrm{i.e.}, at most a Hilbert space with just a few degrees of freedom)
adjoining them via point contacts. In particular, we shall focus on the
important example of tunneling junctions of different types and steady-state conditions.

\section{Case Study: A simple junction out of equilibrium\label{Sec:II}}

How to set up the bosonization formalism so that it remains valid under
nonequilibrium conditions is a delicate procedure, often used but not so very
often discussed in the literature. We shall focus on the important case of
nonequilibrium steady states and we will expand on the details via a
particular example that will also serve to highlight certain subtleties whose
resolution will be the main focus of this paper.

\begin{figure}[ptb]
\begin{center}
\includegraphics[width=\columnwidth]{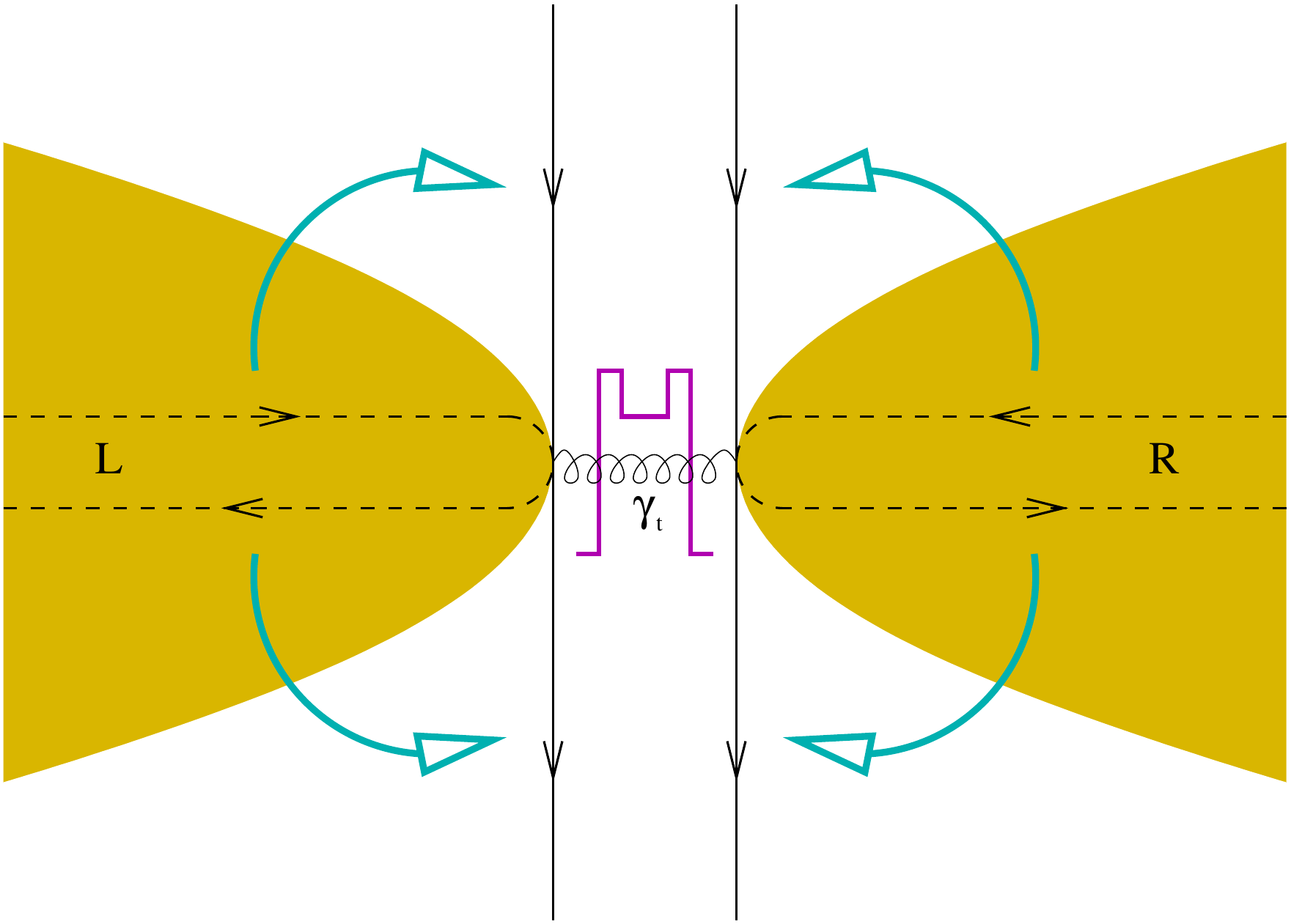}
\end{center}
\caption{Schematic representation of the mapping procedure. When each lead can
be described as fermions on a half line and there are no interactions (or at
least no backscattering), one can unfold the space into a full line
with only one type of movers. These are called chiral fermions and the final
setting is in many ways similar to that which presents itself naturally in
setups involving quantum-Hall edge states. By convention, one sets the
junction at the point $x=x_{0}=0$ (with $\hat{x}$ being the vertical axis in
the figure). In the present case to be studied, the junction is simply a
potential barrier modeled with a tunneling-overlap matrix element
$\gamma_{\text{t}}$. More in general, one could have a more complicated
tunneling system (such as a double barrier with resonant levels in
between) in order to describe tunneling through nanostructures such as quantum
dots. The unfolding procedure can also clearly be generalized to
three-terminal settings and more \cite{shah2006}.}%
\label{Fig_junction}%
\end{figure}

We choose to study the problem of a point-contact junction for noninteracting
spin-1/2 fermions formed between two leads separated by a tunneling barrier.
This situation is captured via a standard tunneling Hamiltonian
\cite{bardeen1961,*cohen1962}. Provided the interactions in the leads are
screened so one can describe them with a noninteracting model, the
problem can be reduced to one dimension via the use of symmetries
\cite{affleck1991} or, alternatively, introducing a lattice regularization and
applying a Lanczos-Haydock recursion \cite{Cini} which is valid even in the
presence of disorder. One ends up with a semi-infinite chain with the junction
or impurity attached to its boundary. In a continuum description (after
introducing an appropriate bandwidth cutoff and linearizing the spectrum) we
have two degrees of freedom which can be called the incoming and outgoing
electrons, that move towards or away from the boundary, respectively. These
are defined for $r>0$ and obey the boundary condition $\psi_{\text{in}%
}^{\dagger}\left(  r=0,t\right)  =\psi_{\text{out}}^{\dagger}\left(
r=0,t\right)  $; we omitted internal indexes for brevity. One can introduce
the new operator $\psi^{\dagger}\left(  r,t\right)  =\theta\left(  -r\right)
\psi_{\text{in}}^{\dagger}\left(  -r,t\right)  +\theta\left(  r\right)
\psi_{\text{out}}^{\dagger}\left(  r,t\right)  $ defined in the whole axis,
where $\theta\left(  r\right)  $ is the Heaviside step function. These are now
chiral fermions and the junction problem has been mapped to two chiral-fermion
leads adjoined at a point (see Fig.~\ref{Fig_junction}). These standard
transformations are commonly used in the context of (quantum-)impurity
problems \cite{affleck1995, affleck2008} (see also Refs.~%
\onlinecite{wilson1975,*bulla2008}
for a related description in the context of numerical renormalization). Notice
that we could equally well exchange $\psi_{\text{in}}^{\dagger}$ and
$\psi_{\text{out}}^{\dagger}$ while defining $\psi^{\dagger}$, which means
that we have the freedom to work with either right or left movers and which
type to choose is a matter of convention. In what follows, after the mapping
to a one-dimensional model, we will refer to the spatial axis as the $\hat{x}$ axis.
The open-circuit (open-junction) boundary conditions now read $\psi^{\dagger
}\left(  x=0^{-},t\right)  =\psi^{\dagger}\left(  x=0^{+},t\right)  $.

\subsection{Setting of the Problem and Direct Solution}

\begin{figure}[ptb]
\begin{center}
\includegraphics[width=\columnwidth]{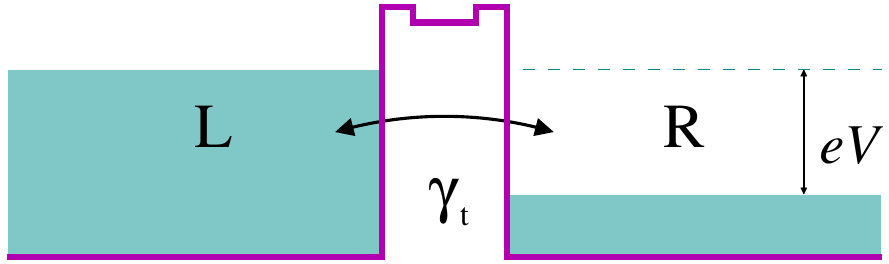}
\end{center}
\caption{Schematic depiction of the setting in which two Fermi seas kept at
different chemical potentials with a difference given by $eV=\mu_{\text{L}%
}-\mu_{\text{R}}$ are connected via quantum tunneling across the potential
barrier that separates them. The tunneling across the classically forbidden
region is modeled by a tight-binding matrix overlap $\gamma_{\text{t}}$ that
can be taken to be energy independent in certain cases (in particular, the
characteristic energy scale of dependence of the transmission coefficient,
$\left\vert \gamma_{\text{t}}\right\vert ^{2}$, has to be much larger than
both $eV$ and $k_{\text{B}}T$ \cite{blanter2000}). Here we assume that the
barrier region does not allow for internal states; a situation when that
happens will be discussed elsewhere \cite{bolech2016}. }%
\label{Fig_barrier}%
\end{figure}

In Hamiltonian language, the model we shall focus on is given by%
\begin{equation}
H=\sum_{\sigma,\ell}\left(  \int\mathcal{H}_{\ell}^{0}\,dx+H_{\text{tun}%
}\right)  \,\text{,}%
\end{equation}
where the Hamiltonian (densities) describing the leads and the Hamiltonian for
the tunneling across a barrier modeled as a quantum point 
contact between the two leads \cite{cuevas1996,*berthod2011}, respectively, are%
\begin{subequations}
\begin{align} 
\mathcal{H}_{\ell}^{0}  &  =v_{\text{F}}\,\psi_{\sigma\ell}^{\dagger}\left(
x,t\right)  \left(  -i\partial_{x}\right)  \psi_{\sigma\ell}\left(  x,t\right)
\\
H_{\text{tun}}  &  =-\gamma_{\text{t}}\,\psi_{\sigma\ell}^{\dagger}\left(
0,t\right)  \psi_{\sigma\bar{\ell}}\left(  0,t\right)\,\mathrm{.}
\end{align}
Here $\psi_{\sigma\ell}\left(  x,t\right)  $ are spin-1/2 ($\sigma
=\uparrow,\downarrow$) chiral fermions in the Heisenberg representation that
are obtained after \textquotedblleft unfolding\textquotedblright\ the two
leads as described above. Notice that the leads are modeled with an exactly
linear dispersion and $v_{\text{F}}$ is the Fermi velocity. The tunneling
matrix element, $\gamma_{\text{t}}$, that characterizes the barrier is taken
to be energy independent and the local fields at $x=0$ are understood in the 
sense of the \textit{local-action} formalism used for the calculations as explained below. 
We took here $\gamma_{\text{t}}$ to be real for
notational simplicity (though that is not necessary and we will write more
general expressions later). Going beyond the standard (physically motivated)
regime, we will consider $\left\vert \gamma_{\text{t}}\right\vert $ to be
arbitrarily large. To fully define the physical situation, we still need to
describe the (nonequilibrium) state of the system. We assume now that at a
much earlier time the connection between the two leads was established and
that there is a battery keeping a constant chemical-potential difference
between the two leads. Let us call $\mu_{\ell}$ the chemical potential of lead
$\ell=$ L$,$R$\,=\mp1$, such that $\mu_{\text{L}}-\mu_{\text{R}}=eV$ and the
full potential drop takes place in the junction region (see
Fig.~\ref{Fig_barrier} for a sketch of the physical configuration). The
information about these chemical potentials will enter into the distribution
functions for each lead. Under these conditions we know, by design, that the
system would have reached some nonequilibrium steady state \cite{doyon2006}
and can be described with a time-translationally invariant action (this is
only important for the particular solutions we discuss below, but not for the
more general conclusions that we reach).

\subsubsection{Transport Characteristics}
\label{Sec:DirectSol}

One of the reasons for defining the problem as we did is that it is amenable
to an exact solution. How to find the transport characteristics is well known
and there exist a number of standard ways of going about it. So we shall be
brief but give a complete summarized account in order to highlight notations
and conventions. Our approach is to integrate out the degrees of freedom in
the leads that are not directly active in the tunneling process (\textrm{i.e.},
away from $x=0$) and thus derive a local action for the problem \cite{Gogolin}.
When doing that, one obtains the diagonal matrix elements as momentum-space
integrals of the two-point Green's functions regularized as principal-value integrals.
Notice this is consistent with the normal-ordering and the regularization of the diagonal 
terms in the action, which are needed for the bosonization treatments that are the 
focus of this work. One can capture the non-equilibrium situation by using a standard 
Schwinger-Keldysh formalism (see Refs.~%
\onlinecite{bolech2004,*bolech2005,*bolech2007}%
~for examples). At the moment we neglect the spin, which will just give a
factor of $2$ at the end, and choose the following spinor basis in Keldysh
space (here we follow the same notation as in Ref.~%
\onlinecite{kakashvili2008a}%
, but we reorder the basis):%
\end{subequations}
\begin{equation}
\Psi=\left(  \psi_{\text{L}}^{\kappa=-}~\psi_{\text{L}}^{\kappa=+}%
~\psi_{\text{R}}^{\kappa=-}~\psi_{\text{R}}^{\kappa=+}\right)  ^{T}%
\end{equation}
where the index $\kappa$ labels the Keldysh-contour branch following the
`minus-means-forward' convention \cite{LandauLifshitzVol10}. Let us define
$\gamma_{\text{t}}=2v_{\text{F}}t$ and use the result for the local inverse
Green's function of the junction \cite{kakashvili2008a},%
\begin{equation}
G^{-1}\left(  \omega\right)  =-2iv_{\text{F}}%
\begin{pmatrix}
-s_{\text{L}} & s_{\text{L}}-1 & it & 0\\
s_{\text{L}}+1 & -s_{\text{L}} & 0 & -it\\
it^{\ast} & 0 & -s_{\text{R}} & s_{\text{R}}-1\\
0 & -it^{\ast} & s_{\text{R}}+1 & -s_{\text{R}}%
\end{pmatrix}
\,\text{,}%
\end{equation}
where $s_{\ell}=s_{\ell}\left(  \omega\right)  \equiv1-2f\left(  \frac
{\omega-\mu_{\ell}}{T_{\ell}}\right)  =\tanh\frac{\omega-\mu_{\ell}}{2T_{\ell
}}$ and $T_{\ell}$ is the temperature of each lead; $f\left(  x\right)
=1/(e^{x}+1)$ is the Fermi function. The current can be computed according to
\begin{align}
\hat{I}  &  =\partial_{t}\frac{\Delta N}{2}=\frac{i}{2}\left[  H,\Delta
N\right]  =\frac{i}{2}\left[  H_{\text{tun}},N_{\text{R}}-N_{\text{L}}\right]
\nonumber\\
&  =i\ell\gamma_{\text{t}}\psi_{\ell}^{\dagger}\left(  0,t\right)  \psi
_{\bar{\ell}}\left(  0,t\right)  \Rightarrow\ell\gamma_{\text{t}}G_{\bar{\ell
}\ell}^{-+}\left(  \delta t=0\right)  \label{Eq:current1}%
\end{align}
(we shall follow the convention in which sums over varying indexes are
implicit). Thus, restoring the complex conjugate tunneling amplitude
$\gamma_{\text{t}}^{\ast}$, we have $I=\left\langle \hat{I}\right\rangle
=-\int\frac{d\omega}{2\pi}\left[  \gamma_{\text{t}}G_{\text{RL}}^{-+}%
-\gamma_{\text{t}}^{\ast}G_{\text{LR}}^{-+}\right]$, where the
expectation value is evaluated via a choice of appropriate
Green's functions in the Schwinger-Keldysh formalism.
Next one can proceed to invert the inverse of the
Green's function and find the necessary expressions for the integrand,%
\begin{equation}
\gamma_{\text{t}}G_{\text{RL}}^{-+}-\gamma_{\text{t}}^{\ast}G_{\text{LR}}%
^{-+}=\frac{2\left\vert t\right\vert ^{2}\left(  s_{\text{L}}-s_{\text{R}%
}\right)  }{\left(  1+\left\vert t\right\vert ^{2}\right)  ^{2}}\,\text{.}%
\end{equation}
More explicitly, and including now the spin-degeneracy factor, we have%
\begin{align}
I  &  =\frac{4\left\vert t\right\vert ^{2}}{\left(  1+\left\vert t\right\vert
^{2}\right)  ^{2}}\int_{-\infty}^{+\infty}\left[  s_{\text{L}}\left(
\omega\right)  -s_{\text{R}}\left(  \omega\right)  \right]  \frac{d\omega
}{2\pi}\label{Eq:IVdirectFT}\\
&  =\frac{4\left\vert t\right\vert ^{2}}{\pi\left(  1+\left\vert t\right\vert
^{2}\right)  ^{2}}\int_{-\infty}^{+\infty}\left[  f_{\text{R}}\left(
\omega\right)  -f_{\text{L}}\left(  \omega\right)  \right]  d\omega
\,\text{.}\nonumber
\end{align}
This expression has a standard form and is intuitively appealing, as the
integrand selects a window (smeared by the temperature) that is $1$ in the
frequency interval between the two chemical potentials and zero outside of it.
The integral can be carried out in general, but we will be taking the
zero-temperature limit ($T_{\text{L}}=T_{\text{R}}=T\rightarrow0$) for
simplicity. In that limit, both $s_{\ell}$ become sign functions ($f_{\ell}$
become step functions) and the integration is trivial:%
\begin{equation}
I\underset{T\rightarrow0}{\longrightarrow}\frac{4\left\vert t\right\vert
^{2}eV}{\pi\left(  1+\left\vert t\right\vert ^{2}\right)  ^{2}}
\label{Eq:IVdirectZT}%
\end{equation}
This gives the particle current, and as always one needs to multiply by
$\left(  -e\right)  $ to get the electric current instead. 
The result is standard \cite{cuevas1996,*berthod2011} and the fact that the
response is exactly linear in $V$ to all orders is a property of the linear
spectrum of the model.

\subsection{Bosonizing in the Steady State}

On the one hand, to study a problem using bosonization, one of the first
things to do is to factor out the fast modes \cite{Giamarchi}. On the other
hand, to study a problem in which a finite voltage bias is present, one of the
first things to do is to introduce it into the calculations (for instance, via
a careful treatment of the interaction picture \cite{mahan2000}). Here, we
need to take care of both things, so it is better to discuss them in the more
formal unified language of gauge transformations. (See, though, Ref.~%
\onlinecite{hershfield1993}
for an approach in terms of scattering states that provides an alternative to
ours but is ultimately equivalent \cite{schiller1998a}.)

As discussed already in the introduction, bosonization is a rewriting of the
excitation spectrum in terms of bosonic degrees of freedom. As such, it does
not capture the information about the reference state or ground state (which
is a Fermi-Dirac sea of noninteracting fermions). Technically, one would say
that what one knows how to bosonize is the normal-ordered Hamiltonian in which
the vev (vacuum expectation value) has been subtracted. The type of
nonequilibrium situation we are considering here presents thus a problem,
because one knows in principle how to normal-order for each lead, but only in
an open-junction configuration. The subsequent inclusion of the tunneling term
constitutes a delicate task. A systematic way of carrying this out starts by
using time-dependent gauge transformations to map the finite-bias problem into
a zero-bias one but with explicitly time-dependent couplings.

Let us first switch to Lagrangian language (to fully capture the effects of a
time-dependent gauge transformation), in which the system is described by the
Lagrangians (densities):%
\begin{subequations}
\begin{align}
\mathcal{L}_{\ell}^{0}  &  =\psi_{\sigma\ell}^{\dagger}\left(  x,t\right)
\left(  i\partial_{t}\right)  \psi_{\sigma\ell}\left(  x,t\right)
-\mathcal{H}_{\ell}^{0}\nonumber\\
&  =\psi_{\sigma\ell}^{\dagger}\left(  x,t\right)  \left(  i\partial
_{t}+iv_{\text{F}}\partial_{x}\right)  \psi_{\sigma\ell}\left(  x,t\right)
\,\text{,}\\
L_{\text{tun}}  &  =-H_{\text{tun}}=\gamma_{\text{t}}\psi_{\sigma\ell
}^{\dagger}\left(  0,t\right)  \psi_{\sigma\bar{\ell}}\left(  0,t\right)
\,\text{.}%
\end{align}
We can now make the following field transformation $\psi_{\sigma\ell}\left(
x,t\right)  =e^{-i\mu_{\ell}t}\tilde{\psi}_{\sigma\ell}\left(  x,t\right)  $,
(notice that some authors follow as an alternative a prescription of including
a time dependence related to the lead chemical potentials into the respective
Klein factors when bosonizing the model \cite{elste2011,*elste2011a}). This
gives%
\end{subequations}
\begin{subequations}
\begin{align}
\mathcal{L}_{\ell}^{0}  &  =\tilde{\psi}_{\sigma\ell}^{\dagger}\left(
x,t\right)  \left(  i\partial_{t}+\mu_{\ell}+iv_{\text{F}}\partial_{x}\right)
\tilde{\psi}_{\sigma\ell}\left(  x,t\right)  \,\text{,}\\
L_{\text{tun}}  &  =e^{i\left(  \mu_{\ell}-\mu_{\bar{\ell}}\right)  t}%
\gamma_{\text{t}}\tilde{\psi}_{\sigma\ell}^{\dagger}\left(  0,t\right)
\tilde{\psi}_{\sigma\bar{\ell}}\left(  0,t\right) \nonumber\\
&  =e^{ieVt}\gamma_{\text{t}}\tilde{\psi}_{\sigma\text{L}}^{\dagger}\left(
0,t\right)  \tilde{\psi}_{\sigma\text{R}}\left(  0,t\right)  +\nonumber\\
&  \qquad\qquad+e^{-ieVt}\gamma_{\text{t}}^{\ast}\tilde{\psi}_{\sigma\text{R}%
}^{\dagger}\left(  0,t\right)  \tilde{\psi}_{\sigma\text{L}}\left(
0,t\right)
\end{align}
(where, in the last line, we restored explicitly the complex conjugate
$\gamma_{\text{t}}^{\ast}$). An important point is that now the distribution
functions in the Keldysh action do not contain information about the chemical
potentials any longer \cite{Altland}. Next we subtract the vev's of each lead
which, by assumption, are the same as those in the absence of the tunneling term
(this is where the Landauer prescription \cite{blanter2000} enters the
calculation) and drop the now ineffectual chemical potential terms. For a
noninteracting problem this is equivalent to factoring out the fast
oscillations in each lead according to an additional field transformation:
$\tilde{\psi}_{\sigma\ell}\left(  x,t\right)  =e^{ik_{\text{F}}^{\ell}x}%
\breve{\psi}_{\sigma\ell}\left(  x,t\right)  $, with $k_{\text{F}}^{\ell}%
=\mu_{\ell}/v_{\text{F}}$ for this linear-dispersion case, and then
subtracting the same (infinite) constant for all leads. So we are naturally
left with the normal-ordered formulation of the problem,%
\end{subequations}
\begin{subequations}
\begin{align}
\mathcal{L}_{\ell}^{0}  &  =\colon\breve{\psi}_{\sigma\ell}^{\dagger}\left(
x,t\right)  \left(  i\partial_{t}+iv_{\text{F}}\partial_{x}\right)
\breve{\psi}_{\sigma\ell}\left(  x,t\right)  \colon\,\text{,}\\
L_{\text{tun}}  &  =e^{ieVt}\gamma_{\text{t}}\breve{\psi}_{\sigma\text{L}%
}^{\dagger}\left(  0,t\right)  \breve{\psi}_{\sigma\text{R}}\left(
0,t\right)  +\nonumber\\
&  \qquad\quad+e^{-ieVt}\gamma_{\text{t}}^{\ast}\breve{\psi}_{\sigma\text{R}%
}^{\dagger}\left(  0,t\right)  \breve{\psi}_{\sigma\text{L}}\left(
0,t\right)  \,\text{.}%
\end{align}
At this point we lost all the information about any absolute-energy reference,
but we still have the information about the potential drop encoded in the
time-dependent phase of the tunneling term (cf.~Fig.~\ref{Fig_barrier}). Given
the infinite-bandwidth setting, we are also in a situation in which space is
naturally to be regarded as half filled. Now one is ready to bosonize the
problem following the standard procedure.

\subsubsection{Abelian Bosonization and Standard Transformations}

The Abelian-bosonization recipe is by now textbook material
\cite{Stone,Gogolin,Giamarchi,shankar1995} and there is no need to present the
details here. There exist though a number of different conventions, which can
bring in some confusion at times. Our notation and conventions follow closely
the review article in Ref.~%
\onlinecite{vondelft1998a}
(which in turn is based on the constructive presentation given earlier by
Haldane \cite{haldane1981}), with the only difference of factors of
$1/\sqrt{2\pi}$ that are needed in order to have a more standard normalization
for the real-space Fermi-field anticommutators \footnote{The less standard
normalization is, however, of spread use in conformal field theory, thus the
proliferation of different conventions.}.

In order to bosonize we go back to the Hamiltonian formulation%
\end{subequations}
\begin{subequations}
\begin{align}
\mathcal{H}_{\ell}^{0}  &  =\colon\breve{\psi}_{\sigma\ell}^{\dagger}\left(
x,t\right)  \left(  -iv_{\text{F}}\partial_{x}\right)  \breve{\psi}%
_{\sigma\ell}\left(  x,t\right)  \colon\,\text{,}\\
H_{\text{tun}}  &  =-e^{ieVt}\gamma_{\text{t}}\breve{\psi}_{\sigma\text{L}%
}^{\dagger}\left(  0,t\right)  \breve{\psi}_{\sigma\text{R}}\left(
0,t\right)  -\nonumber\\
&  \qquad-e^{-ieVt}\gamma_{\text{t}}^{\ast}\breve{\psi}_{\sigma\text{R}%
}^{\dagger}\left(  0,t\right)  \breve{\psi}_{\sigma\text{L}}\left(
0,t\right)  \,\text{,}%
\end{align}
\end{subequations}
and we proceed to bosonize according to $\mathcal{H}_{\ell}^{0}$, which is
akin to working in the interaction picture (with $H_{\text{tun}}$ taken as the
interaction term) \cite{banks1976}. We shall follow the bosonization
prescription \footnote{Recall we have the freedom to work with either
chirality, so we choose to work with `holomorphic' fields; cf.~Ref.~%
\onlinecite{affleck1994}%
.}%
\begin{equation}
\breve{\psi}_{\sigma\ell}\left(  x,t\right)  =\frac{1}{\sqrt{2\pi a}}%
F_{\sigma\ell}\left(  t\right)  e^{-i\phi_{\sigma\ell}\left(  x,t\right)
}\,\text{,}%
\end{equation}
where the $F_{\sigma\ell}\left(  t\right)  $ are the so-called Klein factors
and $a$ is a short-distance regulator \cite{haldane1981}. We shall not include
subleading $1/L$ corrections in the bosonization formulas, because
infinite size is the appropriate limit for a description of the leads in a
Landauer-style transport setup to describe a steady state; as a bonus, this
keeps formulas shorter. In terms of these bosons the Hamiltonian density for
the leads can be shown to take the usual form,%
\begin{equation}
\mathcal{H}^{0}=\sum_{\ell}\mathcal{H}_{\ell}^{0}=\frac{v_{\text{F}}}{4\pi
}\sum_{\sigma=\uparrow,\downarrow;\,\ell=\text{L},\text{R}}\colon\left(
\partial_{x}\phi_{\sigma\ell}\left(  x,t\right)  \right)  ^{2}\colon\,\text{.}%
\end{equation}
One of the main advantages of the bosonic description is that with it one can
more easily recombine degrees of freedom in order to, for instance, separate
the effects of charge and spin dynamics (phenomena such as spin-charge separation
are thus very naturally described with the use of bosonization). Using the
standard, physically motivated, rotated boson basis $\phi_{\sigma\ell}%
=\frac{1}{2}\left(  \phi_{c}+\sigma\phi_{s}+\ell\phi_{l}+\sigma\ell\phi
_{sl}\right)  $, where $\sigma,\ell=\pm1$ when entering as multiplying
factors, the noninteracting Hamiltonian density retains its quadratic form,%
\begin{equation}
\mathcal{H}^{0}=\frac{v_{\text{F}}}{4\pi}\sum_{\nu=c,s,l,sl}\colon\left(
\partial_{x}\phi_{\nu}\left(  x,t\right)  \right)  ^{2}\colon\,\text{,}
\label{Eq:KinEn2}%
\end{equation}
and, as usual, the Klein factors drop out from these terms. We shall refer to
these `physical' sectors as \textit{charge}, \textit{spin}, \textit{lead} (or
flavor), and \textit{spin-lead }(or spin-flavor), respectively. We will see
how they naturally reorganize the information about the physics of tunneling
transport.

Let us now bosonize the tunneling term, rotate the bosons into the physical
sectors, and make some standard simplifications (the sum over $\sigma$ is
implicit and the fields are evaluated at $x=0$):%
\begin{subequations}
\begin{align}
H_{\text{tun}}  &  =-e^{ieVt}\frac{\gamma_{\text{t}}}{2\pi a}F_{\sigma
\text{L}}^{\dagger}F_{\sigma\text{R}}e^{i\phi_{\sigma\text{L}}}e^{-i\phi
_{\sigma\text{R}}}-\nonumber\\
&  \qquad-e^{-ieVt}\frac{\gamma_{\text{t}}^{\ast}}{2\pi a}F_{\sigma\text{R}%
}^{\dagger}F_{\sigma\text{L}}e^{i\phi_{\sigma\text{R}}}e^{-i\phi
_{\sigma\text{L}}}\,\text{,}\label{Eq:bosoHtun1}\\
&  =-e^{ieVt}\frac{\gamma_{\text{t}}}{2\pi a}F_{\sigma\text{L}}^{\dagger
}F_{\sigma\text{R}}e^{-i\left(  \phi_{l}+\sigma\phi_{sl}\right)  }-\nonumber\\
&  \qquad-e^{-ieVt}\frac{\gamma_{\text{t}}^{\ast}}{2\pi a}F_{\sigma\text{R}%
}^{\dagger}F_{\sigma\text{L}}e^{i\left(  \phi_{l}+\sigma\phi_{sl}\right)
}\,\text{.} \label{Eq:bosoHtun2}%
\end{align}
To proceed further we need to take care of the mapping of Klein factors. We
anticipate no subtleties coming from these, but we carry out a careful
treatment nevertheless so as to show that explicitly. The most rigorous way to
proceed is by identifying relations between different bilinears of
\textit{old} and \textit{new} Klein factors, and fixing the four arbitrary
phases that appear
\cite{vondelft1998,*zarand2000,bolech2006a,*iucci2008,rufino2013}:%
\end{subequations}
\begin{subequations}
\begin{align}
F_{\uparrow\text{R}}^{\dagger}F_{\downarrow\text{R}}  &  =F_{sl}^{\dagger
}F_{s}^{\dagger}\,\text{,}\\
F_{\uparrow\text{L}}^{\dagger}F_{\downarrow\text{L}}  &  =F_{sl}F_{s}%
^{\dagger}\,\text{,}\\
F_{\uparrow\text{R}}^{\dagger}F_{\uparrow\text{L}}  &  =F_{sl}^{\dagger}%
F_{l}^{\dagger}\,\text{,}\\
F_{\uparrow\text{R}}^{\dagger}F_{\uparrow\text{L}}^{\dagger}  &
=F_{c}^{\dagger}F_{s}^{\dagger}\,\text{.}%
\end{align}
All the rest of the Klein-factor bilinear relations can be derived from these.
In particular, in order to simplify $H_{\text{tun}}$ we will need the
following ones:%
\begin{align}
F_{\uparrow\text{R}}^{\dagger}F_{\uparrow\text{L}}  &  =F_{sl}^{\dagger}%
F_{l}^{\dagger}\,\text{,}\\
F_{\downarrow\text{R}}^{\dagger}F_{\downarrow\text{L}}  &  =F_{l}^{\dagger
}F_{sl}\,\text{,}\\
F_{\uparrow\text{L}}^{\dagger}F_{\uparrow\text{R}}  &  =F_{l}F_{sl}%
\,\text{,}\\
F_{\downarrow\text{L}}^{\dagger}F_{\downarrow\text{R}}  &  =F_{sl}^{\dagger
}F_{l}\,\text{,}%
\end{align}
(where the last two are simply the Hermitian conjugate of the first two).
Notice that, as one should have expected by looking at the boson fields and
comparing Eqs.~(\ref{Eq:bosoHtun1}) and (\ref{Eq:bosoHtun2}), the right-hand
sides involve only the \textit{lead} and \textit{spin-lead} Klein factors. The
tunneling Hamiltonian density can thus be further rewritten referring only to
the `physical' sectors. One can then undo the steps of the bosonization
procedure and \textit{debosonize} (also called reverse bosonization or
refermionization) in order to arrive again at a problem written in terms of
Fermi fields. Using the standard debosonization prescription, $\breve{\psi
}_{\nu}\left(  x,t\right)  =\frac{1}{\sqrt{2\pi a}}F_{\nu}\left(  t\right)
e^{-i\phi_{\nu}\left(  x,t\right)  }$, which parallels the one we used for
bosonizing in the first place, we arrive at%
\end{subequations}
\begin{subequations}
\begin{align}
\mathcal{H}_{\nu}^{0}  &  =\colon\breve{\psi}_{\nu}^{\dagger}\left(
x,t\right)  \left(  -iv_{\text{F}}\partial_{x}\right)  \breve{\psi}_{\nu
}\left(  x,t\right)  \colon\,\text{,}\\
H_{\text{tun}}  &  =-\left[  e^{ieVt}\gamma_{\text{t}}\breve{\psi}_{l}\left(
0,t\right)  +e^{-ieVt}\gamma_{\text{t}}^{\ast}\breve{\psi}_{l}^{\dagger
}\left(  0,t\right)  \right]  \times\nonumber\\
&  \qquad\qquad\qquad\times\left[  \breve{\psi}_{sl}\left(  0,t\right)
-\breve{\psi}_{sl}^{\dagger}\left(  0,t\right)  \right]  \,\text{.}%
\end{align}
We find that the tunneling term involves only the lead and spin-lead sectors,
while the charge and spin sectors have decoupled from the tunneling process.

The new problem, defined by $H=\int\mathcal{H}^{0}+H_{\text{tun}}$, can now be
regarded as arising from an original problem with the voltage acting as a
chemical-potential shift of the \textit{lead }fermions only ($\nu=l$). In
other words, if we consider the problem given by
\end{subequations}
\begin{subequations}
\begin{align}
\mathcal{H}^{0}  &  =\sum_{\nu}\mathcal{H}_{\nu}^{0}=\psi_{\nu}^{\dagger
}\left(  x,t\right)  \left(  -iv_{\text{F}}\partial_{x}\right)  \psi_{\nu
}\left(  x,t\right)  \,\text{,}\\
H_{\text{tun}}  &  =-\left[  \gamma_{\text{t}}\psi_{l}\left(  0,t\right)
+\gamma_{\text{t}}^{\ast}\psi_{l}^{\dagger}\left(  0,t\right)  \right]
\left[  \psi_{sl}\left(  0,t\right)  -\psi_{sl}^{\dagger}\left(  0,t\right)
\right]
\end{align}
where the chemical potential is set as $\mu_{\nu=l}=-(eV)$ and is zero for all
sectors $\nu\neq l$, this can be connected with the debosonized problem of
interest following equivalent steps to those we presented above via the
combined transformation $\psi_{l}\left(  x,t\right)  =e^{ieV\left(
t-x/v_{\text{F}}\right)  }\breve{\psi}_{l}\left(  x,t\right)  $. Moreover,
this `parent' problem can be seen to be unique (\textrm{i.e.}, there is only
one way to eliminate the time dependence from the tunneling term by
reintroducing chemical potentials into the problem).

\subsection{Indirect Solution using Conventional Bosonization-Debosonization}

One of the goals of a bosonization-debosonization program (BdB for short), as
exemplified above, is to achieve a simplification of the problem at hand that
would not be so easy otherwise. (There could be other alternative or
additional motivations for bosonizing, such as carrying out a
renormalization-group analysis that is more easily done in the bosonic
language; see Ref.~%
\onlinecite{Giamarchi}
for examples.) Indeed, transformations like the one introduced by the simple
rotation of the bosonic basis would be hardly evident if one were to express
them directly in terms of the old and new fermions instead. The example that
we picked is special, because we are able to solve it exactly already in the
original formulation and even in an out-of-equilibrium setting. However, the
BdB program is, in most other cases, crucial for simplifying the problems and
being able to find solutions either exact or approximate.

In the case of our simple junction problem, the BdB program does indeed show
some apparent simplifications. A simple glance at the final form of
$H_{\text{tun}}$ shows that only the \textit{lead} and \textit{spin-lead}
sectors are involved in the transport while the other two sectors
(\textit{spin} and \textit{charge}) do not participate. This provides a
certain economy of description that we will discuss further below. For now,
our immediate goal in this section is to recompute the I-V characteristics of
the junction.

\subsubsection{Re-calculation of Transport after Conventional BdB}
\label{Sec:IndirectSol}

We need again the operator expression of the current, but now in terms of the
new fermionic degrees of freedom. One can translate it from the expression we
gave above [see Eq.~(\ref{Eq:current1})] using BdB or, equivalently, it can be
recomputed directly in terms of the new fields:%
\end{subequations}
\begin{align}
\hat{I}  &  =\partial_{t}\frac{\Delta N}{2}=i\left[  H,\frac{\Delta N}%
{2}\right]  =i\left[  H_{\text{tun}},N_{\nu=l}\right] \\
&  =-i\left[  \psi_{sl}^{\dagger}\left(  0,t\right)  -\psi_{sl}\left(
0,t\right)  \right]  \left[  \gamma_{\text{t}}\psi_{l}\left(  0,t\right)
-\gamma_{\text{t}}^{\ast}\psi_{l}^{\dagger}\left(  0,t\right)  \right]
\,\text{.}\nonumber
\end{align}
Notice that this time the spin degeneracy is already included implicitly in
the formalism. Thus, $I=\left\langle \hat{I}\right\rangle $ is given as%
\begin{align*}
I  &  =-i\gamma_{\text{t}}\left(  \left\langle \psi_{sl}^{\dagger}\left(
0,t\right)  \psi_{l}\left(  0,t\right)  \right\rangle -\left\langle \psi
_{sl}\left(  0,t\right)  \psi_{l}\left(  0,t\right)  \right\rangle \right)
+\\
&  \quad+i\gamma_{\text{t}}^{\ast}\left(  \left\langle \psi_{sl}^{\dagger
}\left(  0,t\right)  \psi_{l}^{\dagger}\left(  0,t\right)  \right\rangle
-\left\langle \psi_{sl}\left(  0,t\right)  \psi_{l}^{\dagger}\left(
0,t\right)  \right\rangle \right)
\end{align*}

Next we calculate the necessary Green's function elements using the same
procedure as in Sec.~\ref{Sec:DirectSol}. However, this time we need to
introduce a Nambu structure due to the presence of \textit{anomalous}
processes in $H_{\text{tun}}$. As a result, we adopt the following spinor
basis (including also the Keldysh indexes and with the frequencies restricted
to the positive semiaxis only in order to avoid double counting):
\begin{widetext}%
\begin{equation}
\Psi\left(  \omega\right)  =%
\begin{pmatrix}
\psi_{l}^{-}\left(  \omega\right)  & \psi_{l}^{+}\left(  \omega\right)  &
\psi_{l}^{\dagger-}\left(  \bar{\omega}\right)  & \psi_{l}^{\dagger+}\left(
\bar{\omega}\right)  & \psi_{sl}^{-}\left(  \omega\right)  & \psi_{sl}%
^{+}\left(  \omega\right)  & \psi_{sl}^{\dagger-}\left(  \bar{\omega}\right)
& \psi_{sl}^{\dagger+}\left(  \bar{\omega}\right)
\end{pmatrix}
^{T}\,\text{.}%
\end{equation}
We write the local inverse Green's function of $H^{0}$, using the fact that
all non-equilibrium Green's functions (\textrm{i.e.}, advanced, retarded, and
Keldysh components) are diagonal in the Nambu basis. The only change required
for the time-reversed Nambu component, as compared with the time-forward one,
is to define $\bar{s}_{\nu}\equiv\tanh\frac{\omega+\mu_{\nu}}{2T_{\nu}}$ for
$\omega$ as given in the argument of the spinor (and we will be taking the
temperature to be uniform, $T_{\nu}=T_{\text{emp}}$). Including also the
contribution of $H_{\text{tun}}$, the local inverse Green's function for the
junction is thus given by%
\begin{equation}
G^{-1}\left(  \omega\right)  =-2iv_{\text{F}}%
\begin{pmatrix}
-s_{l} & s_{l}-1 & 0 & 0 & it^{\ast} & 0 & -it^{\ast} & 0\\
s_{l}+1 & -s_{l} & 0 & 0 & 0 & -it^{\ast} & 0 & it^{\ast}\\
0 & 0 & -\bar{s}_{l} & \bar{s}_{l}-1 & it & 0 & -it & 0\\
0 & 0 & \bar{s}_{l}+1 & -\bar{s}_{l} & 0 & -it & 0 & it\\
it & 0 & it^{\ast} & 0 & -s_{sl} & s_{sl}-1 & 0 & 0\\
0 & -it & 0 & -it^{\ast} & s_{sl}+1 & -s_{sl} & 0 & 0\\
-it & 0 & -it^{\ast} & 0 & 0 & 0 & -\bar{s}_{sl} & \bar{s}_{sl}-1\\
0 & it & 0 & it^{\ast} & 0 & 0 & \bar{s}_{sl}+1 & -\bar{s}_{sl}%
\end{pmatrix}
\,\text{.} \label{Eq:invG2}%
\end{equation}%
\end{widetext}%

\noindent We invert the matrix, identify the relevant matrix elements, and
replace them into the expression for the current. After some algebra one gets%
\begin{equation}
I=\frac{\left\vert t^{\prime}\right\vert ^{2}}{\left(  1+\left\vert t^{\prime
}\right\vert ^{2}\right)  }\int_{0}^{+\infty}\left[  s_{l}\left(
\omega\right)  -\bar{s}_{l}\left(  \omega\right)  \right]  \frac{d\omega}%
{2\pi}\,\text{,}%
\end{equation}
where $t^{\prime}=2t$. The integral can be done in general, but in the
zero-temperature limit reduces to%
\begin{equation}
I\underset{T_{\nu}\rightarrow0}{\longrightarrow}\frac{\left\vert t^{\prime
}\right\vert ^{2}eV}{\pi\left(  1+\left\vert t^{\prime}\right\vert
^{2}\right)  }=\frac{\left(  1+\left\vert t^{\prime}\right\vert ^{2}\right)
}{4}\frac{4\left\vert t^{\prime}\right\vert ^{2}eV}{\pi\left(  1+\left\vert
t^{\prime}\right\vert ^{2}\right)  ^{2}}\,\text{.}%
\end{equation}

We see that the result we obtained for the current shows several discrepancies
from the one in Sec.~\ref{Sec:DirectSol}. Such differences need to be understood.

\section{The Nonequilibrium Transport Puzzle}

We have carefully chosen the nonequilibrium junction problem so that it meets
all the requirements for bosonization to be an exact operator correspondence
between fermions and bosons (cf.~Ref.~%
\onlinecite{vondelft1998a}%
). All the transformations we carried out are thus rigorous and the
discrepancy between the results of Secs.~\ref{Sec:DirectSol} and
\ref{Sec:IndirectSol} is not only unexpected but also unwelcome. There has to
be an inconsistency somewhere and, given that the result of the direct
solution is standard and can be reobtained in a number of alternative ways,
everything seems to indicate that the problem has to be with the indirect
solution. Moreover, the actual transport calculation of the indirect solution
proceeded in a very similar way to the case of the direct one. As a result,
the reason for the discrepancies is likely not in there, but in the preceding
BdB-based mapping used to rewrite the junction problem in terms of the new
fermionic degrees of freedom.

Before furthering the analysis, let us first catalog the discrepancies between
the two solutions:

\begin{enumerate}
\item To match the solutions one needs to arbitrarily correct the tunneling
matrix element of the indirect solution by a factor of $2$ (namely,
$t^{\prime}\mapsto t$) in order to make it look closer to the exact direct solution.

\item There is a overall factor of $4$ difference between the two solutions
(the indirect solution would need to be multiplied by $4$ to match with the
direct solution).

\item There is also an additional factor of $(1+\left\vert t\right\vert ^{2})$
in the numerator of the indirect solution that cancels one power from the
denominator and introduces a further discrepancy with the exact direct solution.
\end{enumerate}

These three discrepancies are present no matter which method we use for the
final transport calculation (they all yield the same result). We highlighted
them by looking at the zero-temperature limit, but it is easy to see that they
are also exactly the same at finite temperature. Additionally, very similar
discrepancies can be seen to be present in equilibrium thermodynamic
calculations using a Matsubara formalism (see the Appendix). Thus, the puzzle
is not restricted only to transport, but it is more evident in transport calculations.

Motivated specially by the third entry from the list of discrepancies, one
could imagine expanding the results of the direct and indirect solutions in
powers of $t$. It is clear that big differences will show up as soon as one
goes beyond leading order in the tunneling matrix element for both
calculations. We therefore expect to be able to gain some insight by studying
the problem using perturbation theory in $t$.

\subsection{A diagrammatic diagnosis}\label{Sec:diagnosis}

Let us start by setting up a dictionary for processes allowed by the different
vertexes in $H_{\text{tun}}$. There are four of those, given by the two
possible spin orientations and the two possible directions of tunneling. Since
our BdB program rests neither on the $SU\left( 2\right)$ invariance nor on
the hermiticity of $H_{\text{tun}}$, we can, in principle, set the four
corresponding matrix elements to different constants and thus individually
trace each process thorough the BdB procedure to construct the dictionary
given in the table below. Alternatively, one can construct the dictionary by
looking at the changes operated by the different \textit{graph vertexes} on
the fermion numbers of the different sectors (which is essentially the
construction that is used to identify the different Klein-factor bilinears
\cite{zarand2000}). The translation between the fermionic structure of the
vertexes in terms of `old' (original) and `new' fermions is thus given by%

\begin{fmffile}{feynart}%
\unitlength= 1mm%
\begin{widetext}%

\begin{center}
\begin{tabularx}{0.55\textwidth}{c|X||c|X}
\hline\noalign{\smallskip}
\multicolumn{4}{c}{\textbf{Simple-junction Graph-vertex Dictionary}}\\
\hline\noalign{\smallskip}
\multicolumn{2}{c||}{\sl Original Fermions} &  \multicolumn{2}{c}{\sl New Fermions} \\
\noalign{\smallskip}\cline{1-4}\noalign{\smallskip}\cline{1-4}\noalign{\smallskip}
$\psi^\dagger_{\uparrow\mathrm{R}}\psi_{\uparrow\mathrm{L}}$ &
\qquad
\begin{fmfgraph*}(20,2)
\fmfleft{i1}
\fmflabel{$\uparrow\mathrm{L}$}{i1}
\fmfright{o1}
\fmflabel{$\uparrow\mathrm{R}$}{o1}
\fmf{fermion}{i1,v1,o1}
\fmfdot{v1}
\end{fmfgraph*}&
$\psi^\dagger_{sl}\psi^\dagger_l$ &
\qquad
\begin{fmfgraph*}(20,2)
\fmfleft{i1}
\fmflabel{$l$}{i1}
\fmfright{o1}
\fmflabel{$sl$}{o1}
\fmf{fermion}{v1,o1}
\fmf{fermion}{v1,i1}
\fmfdot{v1}
\end{fmfgraph*} \\[.5em]
$\psi^\dagger_{\downarrow\mathrm{R}}\psi_{\downarrow\mathrm{L}}$ &
\qquad
\begin{fmfgraph*}(20,2)
\fmfleft{i1}
\fmflabel{$\downarrow\mathrm{L}$}{i1}
\fmfright{o1}
\fmflabel{$\downarrow\mathrm{R}$}{o1}
\fmf{fermion}{i1,v1,o1}
\fmfdot{v1}
\end{fmfgraph*} &
$\psi^\dagger_l\psi_{sl}$ &
\qquad
\begin{fmfgraph*}(20,2)
\fmfleft{i1}
\fmflabel{$sl$}{i1}
\fmfright{o1}
\fmflabel{$l$}{o1}
\fmf{fermion}{i1,v1,o1}
\fmfdot{v1}
\end{fmfgraph*} \\[.5em]
$\psi^\dagger_{\uparrow\mathrm{L}}\psi_{\uparrow\mathrm{R}}$ &
\qquad
\begin{fmfgraph*}(20,2)
\fmfleft{i1}
\fmflabel{$\uparrow\mathrm{R}$}{i1}
\fmfright{o1}
\fmflabel{$\uparrow\mathrm{L}$}{o1}
\fmf{fermion}{i1,v1,o1}
\fmfdot{v1}
\end{fmfgraph*} &
$\psi_l\psi_{sl}$ &
\qquad
\begin{fmfgraph*}(20,2)
\fmfleft{i1}
\fmflabel{$sl$}{i1}
\fmfright{o1}
\fmflabel{$l$}{o1}
\fmf{fermion}{o1,v1}
\fmf{fermion}{i1,v1}
\fmfdot{v1}
\end{fmfgraph*} \\[.5em]
$\psi^\dagger_{\downarrow\mathrm{L}}\psi_{\downarrow\mathrm{R}}$ &
\qquad
\begin{fmfgraph*}(20,2)
\fmfleft{i1}
\fmflabel{$\downarrow\mathrm{R}$}{i1}
\fmfright{o1}
\fmflabel{$\downarrow\mathrm{L}$}{o1}
\fmf{fermion}{i1,v1,o1}
\fmfdot{v1}
\end{fmfgraph*} &
$\psi^\dagger_{sl}\psi_l$ &
\qquad
\begin{fmfgraph*}(20,2)
\fmfleft{i1}
\fmflabel{$l$}{i1}
\fmfright{o1}
\fmflabel{$sl$}{o1}
\fmf{fermion}{i1,v1,o1}
\fmfdot{v1}
\end{fmfgraph*} \\
\noalign{\smallskip}\hline
\end{tabularx}

\end{center}

\noindent Notice that the second two lines are the Hermitian conjugate of the
first two. We can refer to them as (i)-(iv) from top to bottom. Now in order
to calculate the current we need to find the fully dressed vertexes. We can
proceed to dress them by carrying out a perturbative expansion in
$H_{\text{tun}}$ (the Keldysh structure is not important for the present
argument and will be suppressed for the sake of clarity).

Let us consider, for instance, the dressing of vertex (i) in terms of the
original fermions. It proceeds by alternating vertexes (i) and (iii) at
different orders of expansion. Up to third (the first nontrivial) order we
have
\begin{align*}
\qquad
\begin{fmfgraph*}(20,2) \fmfleft{i1} \fmflabel{$\uparrow\mathrm{L}$}{i1} \fmfright{o1} \fmflabel{$\uparrow\mathrm{R}$}{o1} \fmf{fermion}{i1,v1,o1} \fmfblob{.2w}{v1} \end{fmfgraph*}\qquad
=  &  ~\qquad
\begin{fmfgraph*}(20,2) \fmfleft{i1} \fmflabel{$\uparrow\mathrm{L}$}{i1} \fmfright{o1} \fmflabel{$\uparrow\mathrm{R}$}{o1} \fmf{fermion}{i1,v1,o1} \fmfdot{v1} \end{fmfgraph*}\qquad
~+\\
&  +\qquad
\begin{fmfgraph*}(60,2) \fmfleft{i1} \fmflabel{$\uparrow\mathrm{L}$}{i1} \fmfright{o1} \fmflabel{$\uparrow\mathrm{R}$}{o1} \fmf{fermion}{i1,v1} \fmf{fermion,label=$\uparrow\mathrm{R}$}{v1,v2} \fmf{fermion,label=$\uparrow\mathrm{L}$}{v2,v3} \fmf{fermion}{v3,o1} \fmfdot{v1,v2,v3} \end{fmfgraph*}\qquad
~+\\
&  +~\dots
\end{align*}
which in terms of the new fermions translates according to our dictionary
into
\begin{align*}
\qquad
\begin{fmfgraph*}(20,2) \fmfleft{i1} \fmflabel{$l$}{i1} \fmfright{o1} \fmflabel{$sl$}{o1} \fmf{fermion}{v1,i1} \fmf{fermion}{v1,o1} \fmfblob{.2w}{v1} \end{fmfgraph*}\qquad
=  &  ~\qquad
\begin{fmfgraph*}(20,2) \fmfleft{i1} \fmflabel{$l$}{i1} \fmfright{o1} \fmflabel{$sl$}{o1} \fmf{fermion}{v1,i1} \fmf{fermion}{v1,o1} \fmfdot{v1} \end{fmfgraph*}\qquad
~+\\
&  +\qquad
\begin{fmfgraph*}(60,2) \fmfleft{i1} \fmflabel{$l$}{i1} \fmfright{o1} \fmflabel{$sl$}{o1} \fmf{fermion}{v1,i1} \fmf{fermion,label=$sl$}{v1,v2} \fmf{fermion,label=$l$,label.side=left}{v3,v2} \fmf{fermion}{v3,o1} \fmfdot{v1,v2,v3} \end{fmfgraph*}\qquad
~+\\
&  +~\dots
\end{align*}

For spin-down, the diagrams in terms of the original fermions are exactly the
same with the obvious label replacement $(\uparrow)\rightarrow(\downarrow)$.
This corresponds to alternating vertexes (ii) and (iv) at different orders of
expansion. After translation to the new-fermions language one just changes the
labels according to $(\downarrow\mathrm{L})\rightarrow({sl})$ and
$(\downarrow\mathrm{R})\rightarrow({l})$, but this time the arrows of the
fermion propagators stay unchanged (no anomalous processes are involved in
this case, exactly the opposite from the example above with spin-up).

Difficulties arise when we start directly from the new-fermions language and
proceed to dress the vertex in question. This is so because we have additional
(and, we shall claim, unphysical) ways of introducing contractions. Consider,
for instance, again the case of vertex (i). One would proceed to dress it as
follows:
\begin{align*}
\qquad
\begin{fmfgraph*}(20,2) \fmfleft{i1} \fmflabel{$l$}{i1} \fmfright{o1} \fmflabel{$sl$}{o1} \fmf{fermion}{v1,i1} \fmf{fermion}{v1,o1} \fmfblob{.2w}{v1} \end{fmfgraph*}\qquad
=  &  ~ \qquad
\begin{fmfgraph*}(20,2) \fmfleft{i1} \fmflabel{$l$}{i1} \fmfright{o1} \fmflabel{$sl$}{o1} \fmf{fermion}{v1,i1} \fmf{fermion}{v1,o1} \fmfdot{v1} \end{fmfgraph*} \qquad
~+\\
&  + \qquad
\begin{fmfgraph*}(60,2) \fmfleft{i1} \fmflabel{$l$}{i1} \fmfright{o1} \fmflabel{$sl$}{o1} \fmf{fermion}{v1,i1} \fmf{fermion,label=$sl$}{v1,v2} \fmf{fermion,label=$l$,label.side=left}{v3,v2} \fmf{fermion}{v3,o1} \fmfdot{v1,v2,v3} \end{fmfgraph*} \qquad
~+\\[1em]
&  + \qquad
\begin{fmfgraph*}(60,2) \fmfleft{i1} \fmflabel{$l$}{i1} \fmfright{o1} \fmflabel{$sl$}{o1} \fmf{fermion}{v1,i1} \fmf{fermion,label=$sl$,label.side=left}{v2,v1} \fmf{fermion,label=$l$,label.side=left}{v3,v2} \fmf{fermion}{v3,o1} \fmfdot{v1,v2,v3} \end{fmfgraph*} \qquad
~+\\[1em]
&  + \qquad
\begin{fmfgraph*}(60,2) \fmfleft{i1} \fmflabel{$l$}{i1} \fmfright{o1} \fmflabel{$sl$}{o1} \fmf{fermion}{v1,i1} \fmf{fermion,label=$sl$}{v1,v2} \fmf{fermion,label=$l$}{v2,v3} \fmf{fermion}{v3,o1} \fmfdot{v1,v2,v3} \end{fmfgraph*} \qquad
~+\\[1em]
&  + \qquad
\begin{fmfgraph*}(60,2) \fmfleft{i1} \fmflabel{$l$}{i1} \fmfright{o1} \fmflabel{$sl$}{o1} \fmf{fermion}{v1,i1} \fmf{fermion,label=$sl$,label.side=left}{v2,v1} \fmf{fermion,label=$l$}{v2,v3} \fmf{fermion}{v3,o1} \fmfdot{v1,v2,v3} \end{fmfgraph*} \qquad
~+\\
&  + ~\dots
\end{align*}
The four third-order processes correspond to vertex insertions (i-iii-i),
(ii-iv-i), (i-ii-iv) and (ii-i-iv), respectively; which, according to our
dictionary, translated back in terms of the original fermions read as
follows:
\begin{align*}
\qquad
\begin{fmfgraph*}(20,2) \fmfleft{i1} \fmflabel{$\uparrow\mathrm{L}$}{i1} \fmfright{o1} \fmflabel{$\uparrow\mathrm{R}$}{o1} \fmf{fermion}{i1,v1,o1} \fmfblob{.2w}{v1} \end{fmfgraph*}\qquad
=  &  ~ \qquad
\begin{fmfgraph*}(20,2) \fmfleft{i1} \fmflabel{$\uparrow\mathrm{L}$}{i1} \fmfright{o1} \fmflabel{$\uparrow\mathrm{R}$}{o1} \fmf{fermion}{i1,v1,o1} \fmfdot{v1} \end{fmfgraph*} \qquad
~+\\
&  + \qquad
\begin{fmfgraph*}(60,2) \fmfleft{i1} \fmflabel{$\uparrow\mathrm{L}$}{i1} \fmfright{o1} \fmflabel{$\uparrow\mathrm{R}$}{o1} \fmf{fermion}{i1,v1} \fmf{fermion,label=$\uparrow\mathrm{R}$}{v1,v2} \fmf{fermion,label=$\uparrow\mathrm{L}$}{v2,v3} \fmf{fermion}{v3,o1} \fmfdot{v1,v2,v3} \end{fmfgraph*} \qquad
~+\\[1em]
&  + \qquad
\begin{fmfgraph*}(60,2) \fmfleft{i1} \fmflabel{$\downarrow\mathrm{L}$}{i1} \fmfright{o1} \fmflabel{$\uparrow\mathrm{R}$}{o1} \fmf{fermion}{i1,v1} \fmf{fermion,label=$\downarrow\mathrm{R}$}{v1,v2} \fmf{fermion,label=$\downarrow\mathrm{L}\mid\;\uparrow\mathrm{L}$}{v2,v3} \fmf{fermion}{v3,o1} \fmfdot{v1,v2,v3} \end{fmfgraph*} \qquad
~+\\[1em]
&  + \qquad
\begin{fmfgraph*}(60,2) \fmfleft{i1} \fmflabel{$\uparrow\mathrm{L}$}{i1} \fmfright{o1} \fmflabel{$\downarrow\mathrm{L}$}{o1} \fmf{fermion}{i1,v1} 
\fmf{fermion,label=$\uparrow\mathrm{R}\mid\;\downarrow\mathrm{L}$}{v1,v2} \fmf{fermion,label=$\downarrow\mathrm{R}$}{v2,v3} \fmf{fermion}{v3,o1} \fmfdot{v1,v2,v3} \end{fmfgraph*} \qquad
~+\\[1em]
&  + \qquad
\begin{fmfgraph*}(60,2) \fmfleft{i1} \fmflabel{$\downarrow\mathrm{L}$}{i1} \fmfright{o1} \fmflabel{$\downarrow\mathrm{L}$}{o1} \fmf{fermion}{i1,v1} 
\fmf{fermion,label=$\downarrow\mathrm{R}\mid\;\uparrow\mathrm{L}$}{v1,v2} \fmf{fermion,label=$\uparrow\mathrm{R}\mid\;\downarrow\mathrm{R}$}{v2,v3} \fmf{fermion}{v3,o1} \fmfdot{v1,v2,v3} \end{fmfgraph*} \qquad
~+\\
&  + ~\dots
\end{align*}
\end{widetext}%
\end{fmffile}%

The last three contractions are not allowed in the original-fermions
framework, as they require spin flip and some even $\mathrm{L}\leftrightarrow
\mathrm{R}$ exchange (as indicated by the inner labelings). Moreover, they do
not even dress the correct vertex (as indicated by the outer labelings). From
a practical point of view, one may notice that while we deal with four
distinct types of original fermions ($\uparrow\mathrm{L}$, $\downarrow
\mathrm{L}$, $\uparrow\mathrm{R}$ and $\downarrow\mathrm{R}$), we deal with
only two types of new fermions ($l$ and $sl$ ). We conclude that the more
compact description achieved after the BdB-based mapping introduces the
possibility of spurious processes that should not have been there. These are
processes that mix vertexes (i) and (iii) with vertexes (ii) and (iv), which
in terms of the original fermions cannot happen due to spin conservation. This
clearly hints at the possibility that, in the new-fermions framework, the
\textit{spin} sector should not really be decoupled after all.

\section{Consistent Approach to Bosonization-Debosonization}

We need to revisit the transformations in the BdB-based mapping used above,
with the goal of finding the source of the discrepancies with respect to the
direct calculations. In particular, one needs to be careful about the fact
that the tunneling term is not normal ordered (since the procedure of
subtracting the vev is not well defined for the processes in $H_{\text{tun}}$
for they are not diagonal in fermion `internal indexes').

\subsection{Keys to Consistency}

We proceed to study again the bosonization of the tunneling term but taking
care of \textit{not combining exponentials}. If we start from
Eq.~(\ref{Eq:bosoHtun1}) and perform the change of basis for the bosons, we
arrive at%
\begin{align*}
H_{\text{tun}}  &  =-e^{ieVt}\frac{\gamma_{\text{t}}}{2\pi a}F_{\sigma
\text{L}}^{\dagger}F_{\sigma\text{R}}e^{i\left(  \phi_{c}+\sigma\phi_{s}%
-\phi_{l}-\sigma\phi_{sl}\right)  /2}\times\\
&  \qquad\qquad\qquad\qquad\qquad\times e^{-i\left(  \phi_{c}+\sigma\phi
_{s}+\phi_{l}+\sigma\phi_{sl}\right)  /2}-\\
&  \quad-e^{-ieVt}\frac{\gamma_{\text{t}}^{\ast}}{2\pi a}F_{\sigma\text{R}%
}^{\dagger}F_{\sigma\text{L}}e^{i\left(  \phi_{c}+\sigma\phi_{s}+\phi
_{l}+\sigma\phi_{sl}\right)  /2}\times\\
&  \qquad\qquad\qquad\qquad\qquad\times e^{-i\left(  \phi_{c}+\sigma\phi
_{s}-\phi_{l}-\sigma\phi_{sl}\right)  /2}%
\end{align*}
We will now, on the one hand, combine the exponentials in which the bosons
appear with the same sign (we are prompted to do this by a study of the
corresponding operator product expansions, OPEs, and by the consistency with
the mapping of the Klein factors
\cite{vondelft1998,*zarand2000,bolech2006a,*iucci2008}). On the other hand, we
will be careful not to combine the exponentials in which the bosonic exponents
appear with opposite signs (prompted by the suspicion, from our perturbative
analysis, that the $\nu=c,s$ sectors should not completely decouple from the
tunneling process). We will discuss the \textit{charge} and \textit{spin}
sectors carefully momentarily; for now we debosonize in the \textit{lead} and
\textit{spin-lead} sectors only (using the same prescription that was
introduced above). The tunneling term takes the following form (all the fields
are evaluated at $x=0$ and at time $t$):%
\begin{align*}
H_{\text{tun}}  &  =-e^{ieVt}\gamma_{\text{t}}\breve{\psi}_{l}\breve{\psi
}_{sl}e^{i\phi_{c}/2}e^{-i\phi_{c}/2}e^{i\phi_{s}/2}e^{-i\phi_{s}/2}-\\
&  \quad-e^{ieVt}\gamma_{\text{t}}\breve{\psi}_{sl}^{\dagger}\breve{\psi}%
_{l}e^{i\phi_{c}/2}e^{-i\phi_{c}/2}e^{-i\phi_{s}/2}e^{i\phi_{s}/2}-\\
&  \quad-e^{-ieVt}\gamma_{\text{t}}^{\ast}\breve{\psi}_{sl}^{\dagger}%
\breve{\psi}_{l}^{\dagger}e^{i\phi_{c}/2}e^{-i\phi_{c}/2}e^{i\phi_{s}%
/2}e^{-i\phi_{s}/2}-\\
&  \quad-e^{-ieVt}\gamma_{\text{t}}^{\ast}\breve{\psi}_{l}^{\dagger}%
\breve{\psi}_{sl}e^{i\phi_{c}/2}e^{-i\phi_{c}/2}e^{-i\phi_{s}/2}e^{i\phi
_{s}/2}\,\text{;}%
\end{align*}
which is the same as before but with the addition of the extra exponential factors.

A pragmatic way to proceed in order to debosonize in the \textit{charge} and
\textit{spin} sectors as well is by replacing the vertex products by
lattice-like fermionic densities according to the prescription%
\begin{equation}
e^{\pm i\phi_{c,s}/2}e^{\mp i\phi_{c,s}/2}\mapsto\tilde{n}_{c,s}^{\pm
}\,\text{.}%
\end{equation}
These new objects (to be defined and discussed more in detail below) can be
interpreted as particle and hole densities for new fermionic degrees of
freedom in the \textit{charge} and \textit{spin} sectors. They shall be
considered in their `eigen-basis' and they have eigenvalues $0$ or $1$ and $1$
or $0$, respectively and correspondingly. This is the central result of the
\textit{consistent} way to debosonize and the (almost) final form of the
tunneling Hamiltonian is%
\begin{align*}
H_{\text{tun}}  &  =-e^{ieVt}\gamma_{\text{t}}\tilde{n}_{c}^{+}\tilde{n}%
_{s}^{+}\breve{\psi}_{l}\breve{\psi}_{sl}-e^{ieVt}\gamma_{\text{t}}\tilde
{n}_{c}^{+}\tilde{n}_{s}^{-}\breve{\psi}_{sl}^{\dagger}\breve{\psi}_{l}-\\
&  \quad-e^{-ieVt}\gamma_{\text{t}}^{\ast}\tilde{n}_{c}^{+}\tilde{n}_{s}%
^{+}\breve{\psi}_{sl}^{\dagger}\breve{\psi}_{l}^{\dagger}-e^{-ieVt}%
\gamma_{\text{t}}^{\ast}\tilde{n}_{c}^{+}\tilde{n}_{s}^{-}\breve{\psi}%
_{l}^{\dagger}\breve{\psi}_{sl}\,\text{.}%
\end{align*}
It can be easily seen that the inclusion of the $\tilde{n}$ factors naturally
avoids the mixing of graph-vertexes (i) and (iii) with graph-vertexes (ii) and
(iv), exactly as was concluded to be necessary in the diagrammatic
discussion of the previous section (Sec.~\ref{Sec:diagnosis}). 
Notice also that, in the same vein, these
factors also stop us from being able to rewrite $H_{\text{tun}}$ in terms of
Majorana-fermion combinations.

In the next two subsections we provide some additional rationale, but those
readers that want to skip some of the technical discussion can jump ahead to
the last subsection of this section (Sec.~\ref{Sec:Resolution}) and see how we
are now able to recover exactly the results of the direct calculation (which
can be taken as a pragmatic justification for the procedure).

\subsection{Matters of Regularization}

The exponentials of bosonic fields of the type $e^{i\lambda\phi_{\nu}}$ are
central objects in the bosonization formalism known as \textit{vertex
operators}. The bosonization prescription tells us that $\psi_{\nu}^{\dagger
}\propto e^{i\phi_{\nu}}$ (with $\lambda=1$) while normal-ordered densities
are bosonized according to $\colon\psi_{\nu}^{\dagger}\psi_{\nu}\colon
=\frac{1}{2\pi}\partial\phi_{\nu}$. The consistency between these two
prescriptions can be checked by bosonizing the non-normal-ordered case,
$\psi_{\nu}^{\dagger}\psi_{\nu}=e^{i\phi_{\nu}}e^{-i\phi_{\nu}}/2\pi a$, and
expanding the right-hand side by using known results for the OPEs of vertex
operators \cite{vondelft1998a}.

However, when bosonizing, oftentimes our aim is to change basis from the
spin-\&-lead (or spin-\&-flavor) states to a basis that separates physical
sectors (\textit{charge}, \textit{spin}, \textit{lead} and \textit{spin-lead})
because some of the physics will simplify by doing that (this is the
transformation that we performed in $H_{\text{tun}}$). Proceeding formally for
each vertex operator of a density operator [using $\phi_{\sigma\ell}\left(
x\right)  =\sum_{n}\phi_{\nu_{n}}/2$, where the $\nu_{n}$ label the physical
sectors and we absorbed minus signs that are not important for this part of
the discussion], we have%
\begin{subequations}
\begin{align}
e^{i\phi_{\sigma\ell}}e^{-i\phi_{\sigma\ell}}  &  =\prod\limits_{n}%
e^{i\phi_{\nu_{n}}/2}e^{-i\phi_{\nu_{n}}/2}\label{Eq:SBC}\\
\left[  1+a\,\partial\phi_{\sigma\ell}+\ldots\right]   &  \approx1+\sum
_{n}a\,\partial\phi_{\nu_{n}}/2+\ldots\\
1+\delta n_{\sigma\ell}+\ldots &  \approx1+\sum_{n}\delta n_{\nu_{n}}/2+\ldots
\end{align}
where in the third line we introduced lattice-like density fluctuations,
$\delta n_{\sigma\ell}\equiv a\,\partial\phi_{\sigma\ell}$ and $\delta
n_{\nu_{n}}\equiv a\,\partial\phi_{\nu_{n}}$, to stress that they need to be
small in order to connect to the first line. (A standard view is to treat $a$
as a control parameter for the expansions in the second line.) Therefore,
these transformations are consistent if bosonization is treated as an
expansion around a half-filled ground state (in a real-space picture). While
the bosonization identities are precise, some manipulations might not hold
when the deviations from the local half-filled state are large. If a
particular problem, as is the case of some transport problems like the one
that we are studying, forces us to consider large $\delta n$ fluctuations,
then we need to proceed with caution while expanding.

One solution is to expand around a different state, which can be achieved via
a linear transformation. Consider the following vertex OPE at some $x=x_{0}$
(the position of the junction or impurity) and treat $a$ as an expansion
parameter (not necessarily small) \footnote{From this OPE it also follows that
$\left\{  e^{i\phi_{\sigma\ell}},e^{-i\phi_{\sigma\ell}}\right\}  \approx2$.
Thus $e^{\pm i\phi_{\sigma\ell}}/\sqrt{2}$ would be properly normalized to be
regarded as fermionic ladder operators.}:%
\end{subequations}
\[
e^{i\phi_{\sigma\ell}}e^{-i\phi_{\sigma\ell}}\approx1+a\,\partial\phi
_{\sigma\ell}+\ldots\equiv1+\left(  1+a\,\partial\tilde{\phi}_{\sigma\ell
}\right)  +\ldots
\]
This serves as a definition of a \textit{shifted} set of bosons, $\tilde{\phi
}_{\sigma\ell}$, which are used to expand around a differently filled state
(unit-filling in this case) and need to obey $\partial\tilde{\phi}_{\sigma
\ell}=\partial\phi_{\sigma\ell}-1/a$. Reintroducing the $x$ dependence from
the OPE before taking the $a\rightarrow0$ limit [\textrm{i.e.}, replacing
$1/a\mapsto\pi\delta\left(  x-x_{0}\right)  $],\ and integrating this relation
one gets $\tilde{\phi}_{\sigma\ell}\left(  x\right)  =\phi_{\sigma\ell}\left(
x\right)  -\frac{\pi}{2}\operatorname{sgn}\left(  x-x_{0}\right)  $, up to an
additive constant. The new bosons have identical commutation relations and
OPEs except at $x=x_{0}$ due to the presence of these solitonic shifts.

Expanding around $a\partial\phi_{\sigma\ell}=1$ is equivalent to expanding
around $a\partial\tilde{\phi}_{\sigma\ell}=0$ and we can use small-variable
expansions in terms of the latter. For the kind of vertex products we are
considering (at $x=x_{0}$) we have%
\begin{subequations}
\begin{align}
e^{i\phi_{\sigma\ell}}e^{-i\phi_{\sigma\ell}}  &  \approx2\left(  1+\frac
{a}{2}\partial\tilde{\phi}_{\sigma\ell}+\ldots\right)  +\ldots\\
&  \approx2\sqrt{1+a\partial\tilde{\phi}_{\sigma\ell}}+\ldots\label{Eq:resum}%
\\
&  \approx2\sqrt{e^{i\tilde{\phi}_{\sigma\ell}}e^{-i\tilde{\phi}_{\sigma\ell}%
}}\,\text{;}%
\end{align}
\end{subequations}
where to get to the second line we used a Taylor expansion for the square root
of a binomial (the first two lines are strictly equivalent to the order that
is given explicitly; their connection can be regarded as a sort of partial
re-summation that is also consistent with a further study of other vertex OPEs
that we carried out as well). Alternatively, applying the same vertex-vertex
OPE, but in reverse, to the parentheses in the first line of the equation
above we have%
\begin{equation}
e^{i\phi_{\sigma\ell}}e^{-i\phi_{\sigma\ell}}\approx2\,e^{i\tilde{\phi
}_{\sigma\ell}/2}e^{-i\tilde{\phi}_{\sigma\ell}/2}\,\text{.}%
\end{equation}
This implies that, generically,%
\begin{equation}
e^{i\phi/2}e^{-i\phi/2}\approx\sqrt{e^{i\phi}e^{-i\phi}}\,\text{;}
\label{Eq:SqrRoot}%
\end{equation}
as will be proven below by working to all orders without resorting to OPEs
(see Eq.~\ref{Eq:n2_eq_n}).

There is a delicate point regarding the proper normalization (or scaling of
the coupling constants) of non-normal-ordered terms as those in $H_{\text{tun}%
}$. This is more easily understood considering vertex products diagonal in
internal indexes. Using the consistent identities derived above, we can
proceed as follows:%
\begin{subequations}
\begin{align}
e^{i\phi_{\sigma\ell}}e^{-i\phi_{\sigma\ell}}  &  \approx2\sqrt{e^{i\tilde
{\phi}_{\sigma\ell}}e^{-i\tilde{\phi}_{\sigma\ell}}}\\
&  \approx\frac{1}{2}\sqrt{16\prod\limits_{n}e^{i\tilde{\phi}_{\nu_{n}}%
/2}e^{-i\tilde{\phi}_{\nu_{n}}/2}}\\
&  \approx\frac{1}{2}\sqrt{\prod\limits_{n}2\sqrt{e^{i\tilde{\phi}_{\nu_{n}}%
}e^{-i\tilde{\phi}_{\nu_{n}}}}}\\
&  \approx\frac{1}{2}\prod\limits_{n}\sqrt{e^{i\phi_{\nu_{n}}^{\prime}%
}e^{-i\phi_{\nu_{n}}^{\prime}}}\\
&  \approx\frac{1}{2}\prod\limits_{n}e^{i\phi_{\nu_{n}}^{\prime}/2}%
e^{-i\phi_{\nu_{n}}^{\prime}/2}\,\text{,} \label{Eq:CBC}%
\end{align}
where (i) in the first line we shifted the bosons away from half filling; (ii)
from the first to the second line we did a change of basis; (iii) going to the
third line we used Eq.~(\ref{Eq:SqrRoot}); (iv) from the third to the fourth
line we shifted the bosons back to half filling and we also distributed the
overall square root; and (v) finally we redistributed the square root between
the two vertex operators again using Eq.~(\ref{Eq:SqrRoot}). Notice the
introduction of the primes (in $\phi_{\nu}^{\prime}$) to distinguish this case
when the change of basis is done with the $\tilde{\phi}$'s from the case when
it was done directly with the original $\phi$'s. The primes will be dropped
when a comparison is not being done and the case in point is clear from the
context (this notational variation is used in this subsection only).

What we found is that if the change of bosonic basis is done in terms of
\textit{shifted bosons}, then a prefactor of $1/2$ appears for proper
normalization (and we shall make this conclusion extensive to nondiagonal
products as well). This kind of normalization changes, or rescaling of
couplings, is common in bosonization treatments and can often be traced to
subtle differences in regularization schemes. In particular, shifting the
bosons is equivalent to acting with so-called \textit{boundary-condition
changing operators} \cite{Gogolin,schotte1969,*affleck1994,*ye1997a,*shah2003}%
, which are a known source for \textquotedblleft coupling-constant
redefinitions\textquotedblright\ (for another example, also involving a
relative factor of $2$, the reader can look at Sec.~2 of Appendix A in Ref.~%
\onlinecite{zarand2000}%
). To summarize our result, we should contrast the differences between
Eq.~(\ref{Eq:CBC}) and the one we presented at the start of this subsection in
Eq.~(\ref{Eq:SBC}). To develop some intuition, let us introduce the
lattice-like notation $2n_{\nu}\approx1+\delta n_{\nu}$, so that, near
half filling, the left-hand side is close to $1$ and near maximum filling it is close to
$2$. We will also use the notation $\sqrt{2}\tilde{n}_{\nu}\approx
\sqrt{2n_{\nu}}\approx\sqrt{1+\delta n_{\nu}}$ (these will be made more
precise in the next subsection). The two BdB-mapping relations can then be
rewritten as%
\end{subequations}
\begin{subequations}
\begin{align}
2n_{\sigma\ell}  &  \approx\prod\limits_{n}\sqrt{2}\tilde{n}_{\nu_{n}}%
\quad\text{when}\quad2n_{\sigma\ell}\approx1\,\text{,}\\
2n_{\sigma\ell}  &  \approx\frac{1}{2}\prod\limits_{n}\sqrt{2}\tilde{n}%
_{\nu_{n}}\quad\text{when}\quad2n_{\sigma\ell}\approx2~\text{(or
}0\text{)\thinspace;}%
\end{align}
where we highlighted that they are useful in different regimes. Which one, or
when each of the two, should be used needs to be judged depending on the
problem that is being solved (and that is part of what we mean by a
\textit{consistent} use of BdB-based transformations). We argue that the
junction problem requires the use of the second one, because the physics
of tunneling calls for the consideration of unit-size particle-number fluctuations 
at the junction ($n_{\sigma\ell}=0\leftrightarrow1$).

It is instructive to see how these two different regimes (\textrm{i.e.},
half filling \textrm{versus} maximum/minimum filling) are connected in our formalism
to a change of boundary conditions for the new fermions after the BdB-based
mapping. We start from the continuum boundary conditions, $\psi_{\sigma\ell
}^{\dagger}\left(  0^{-}\right)  =\psi_{\sigma\ell}^{\dagger}\left(
0^{+}\right)  $ for all $\sigma\ell$. After changing basis in the intermediate
bosonic language of the $\phi$'s, we get to the new fields with $\psi_{\nu
_{n}}^{\dagger}\left(  0^{-}\right)  =\psi_{\nu_{n}}^{\dagger}\left(
0^{+}\right)  $ for all $\nu_{n}=c,s,l,sl\,$, as naturally expected. If we do
the change of basis with the $\tilde{\phi}$'s instead, the resulting boundary
conditions are different. In the charge sector, from the definition of the
$\tilde{\phi}$'s it follows that $\tilde{\phi}_{c}=\phi_{c}-\pi
\operatorname{sgn}\left(  x\right)  $, (where we went back to $x_{0}=0$), and
thus $\phi_{c}^{\prime}=\tilde{\phi}_{c}+\frac{\pi}{2}\operatorname{sgn}%
\left(  x\right)  =\phi_{c}-\frac{\pi}{2}\operatorname{sgn}\left(  x\right)$. 
From there it follows that $\psi_{c}^{\dagger}\left(  0^{-}\right)
=-\,\psi_{c}^{\dagger}\left(  0^{+}\right)  $. For the other sectors ($\nu
_{n}\neq c$), we simply have $\tilde{\phi}_{\nu_{n}}=\phi_{\nu_{n}}$, and thus
$\phi_{\nu_{n}}^{\prime}=\tilde{\phi}_{\nu_{n}}+\frac{\pi}{2}%
\operatorname{sgn}\left(  x\right)  =\phi_{\nu_{n}}+\frac{\pi}{2}%
\operatorname{sgn}\left(  x\right)$; so that, in a different way, we still
get that $\psi_{\nu_{n}}^{\dagger}\left(  0^{-}\right)  =-\,\psi_{\nu_{n}%
}^{\dagger}\left(  0^{+}\right)  $.

Remarkably, these antiperiodic boundary conditions parallel what Affleck calls
\textquotedblleft strong-coupling boundary conditions\textquotedblright\ in
the context of the boundary-conformal-field-theory approach to
quantum-impurity problems [see Eq.~(1.29) of Ref.~%
\onlinecite{affleck2008}%
]. The name is because these are the type of boundary conditions needed in the
strong-coupling limit of those problems. What these boundary conditions
actually do is to decouple the band-fermion degrees of freedom at $x=x_{0}$
from the rest of the bulk; that way they are not tied to half filling (or
other) conditions and they are available to couple them (strongly) to the
impurity. In our case, we shall in general need those degrees of freedom to be
available (even if there is no impurity) to participate unrestrainedly in
transport situations.

We shall thus refer in our context more generically to \textit{consistent
boundary conditions} (CBCs). These depend on the problem at hand and in the
particular example studied here they turn out to be antiperiodic boundary
conditions. Notice that the need for a factor of $1/2$ as discussed above can
be seen as the practical manifestation of the boundary conditions that were
(implicitly) adopted. Let us also mention that the use of CBCs does not modify
the form of the kinetic part of the action (\textrm{i.e.}~when rewriting
Eq.~\ref{Eq:KinEn2} in terms of $\phi\rightarrow\phi^{\prime}$). The solitons
that we introduce with $\phi^{\prime}$ will induce in $\mathcal{H}^{0}$
additional slips of $2\pi$ localized to a length scale of $a$ around $x_{0}$,
but since that is the limit of length resolution and the bosonic fields are
compact with radius $2\pi$, those contributions consistently drop out.

\subsection{Tunneling of New Fermions}

It is now a matter of a delicate but ultimately simple replacement to finish
the debosonization of $H_{\text{tun}}$ in the \textit{charge} and
\textit{spin} sectors. We (re)introduce the following definitions (all fields
are at $x_{0}=0$ and time $t$):%
\end{subequations}
\begin{subequations}
\begin{align}
\sqrt{2}\tilde{n}_{c,s}^{+}  &  \equiv e^{i\phi_{c,s}/2}e^{-i\phi_{c,s}/2}\\
&  \equiv\sqrt{1+a\,\partial\phi_{c,s}}=\sqrt{e^{i\phi_{c,s}}e^{-i\phi_{c,s}}%
}\nonumber\\
&  \equiv\sqrt{2\pi a~\psi_{c,s}^{\dagger}\psi_{c,s}}=\sqrt{2n_{c,s}^{+}%
}\,\text{,}\nonumber\\
&  ~\nonumber\\
\sqrt{2}\tilde{n}_{s}^{-}  &  \equiv e^{-i\phi_{s}/2}e^{i\phi_{s}/2}\\
&  \equiv\sqrt{1-a\,\partial\phi_{c,s}}=\sqrt{e^{-i\phi_{s}}e^{i\phi_{s}}%
}\nonumber\\
&  \equiv\sqrt{2\pi a~\psi_{s}\psi_{s}^{\dagger}}=\sqrt{2n_{s}^{-}}%
\,\text{.}\nonumber
\end{align}
Thus $\tilde{n}_{c,s}^{+}$ ($\tilde{n}_{s}^{-}$) are simply the square
roots of the particle (hole) density of \textit{charge} or \textit{spin}
fermions `\textrm{at the site}' of the junction. For a physical picture, one
could think of them as corresponding to a single lattice site after a lattice
discretization with $\pi a$ as the lattice constant; even though that is
\textit{not} the type of regularization adopted when bosonizing
\cite{haldane1981}. One can explicitly check consistency by calculating their
squares via the equal-time, \textit{full} operator product. For that we need
to write the bosons in terms of their creation and annihilation components,
$\phi\left(  x\right)  =\varphi^{\dagger}\left(  x\right)  +\varphi\left(
x\right)  ,$ which obey $\left[  \varphi\left(  x\right)  ,\varphi^{\dagger
}\left(  x^{\prime}\right)  \right]  =-\ln\left(  1-e^{-\frac{2\pi}{L}\left[
i\left(  x-x^{\prime}\right)  +a\right]  }\right)  $; (see Ref.~%
\onlinecite{vondelft1998a}
for the notational convention to point-split the product and normal-order the
vertex operators):
\end{subequations}
\begin{align}
\left[  \tilde{n}\right]  ^{2}  &  =\frac{1}{2}e^{i\phi\left(  x\right)
/2}e^{-i\phi\left(  x^{\prime}\right)  /2}e^{i\phi\left(  x\right)
/2}e^{-i\phi\left(  x^{\prime}\right)  /2}\nonumber\\
&  =\sqrt{\frac{\pi a}{2L}}e^{\frac{i}{2}\varphi^{\dagger}\left(  x\right)
}e^{\frac{i}{2}\varphi\left(  x\right)  }e^{-\frac{i}{2}\varphi^{\dagger
}\left(  x^{\prime}\right)  }e^{-\frac{i}{2}\varphi\left(  x^{\prime}\right)
}\times\nonumber\\
&  \quad\times e^{\frac{i}{2}\varphi^{\dagger}\left(  x\right)  }e^{\frac
{i}{2}\varphi\left(  x\right)  }e^{-\frac{i}{2}\varphi^{\dagger}\left(
x^{\prime}\right)  }e^{-\frac{i}{2}\varphi\left(  x^{\prime}\right)
}\nonumber\\
&  =\sqrt{\frac{\pi a}{2L}}\left(  \frac{1-e^{-\frac{2\pi}{L}\left[  i\left(
x-x^{\prime}\right)  +a\right]  }}{1-e^{-\frac{2\pi}{L}\left[  i\left(
x^{\prime}-x\right)  +a\right]  }}\right)  ^{1/4}\times\nonumber\\
&  \quad\times\sqrt{1-e^{-\frac{2\pi}{L}a}}\,e^{i\varphi^{\dagger}\left(
x\right)  }e^{i\varphi\left(  x\right)  }e^{-i\varphi^{\dagger}\left(
x^{\prime}\right)  }e^{-i\varphi\left(  x^{\prime}\right)  }\nonumber\\
&  \approx\frac{\pi a}{L}e^{i\varphi^{\dagger}\left(  x\right)  }%
e^{i\varphi\left(  x\right)  }e^{-i\varphi^{\dagger}\left(  x^{\prime}\right)
}e^{-i\varphi\left(  x^{\prime}\right)  }\nonumber\\
&  =\frac{1}{2}e^{i\phi\left(  x\right)  }e^{-i\phi\left(  x^{\prime}\right)
}\nonumber\\
&  =n\,\text{,} \label{Eq:n2_eq_n}%
\end{align}
where $\phi$ stands for either $\pm\phi_{c}$ or $\pm\phi_{s}$.

The squares of the $\tilde{n}$'s have the properties that $\left(
n_{c,s}^{\pm}\right)  ^{2}=n_{c,s}^{\pm}$ and $n_{c,s}^{\pm}n_{c,s}^{\mp}=0$;
we shall refer to these as \textit{idempotence} and \textit{co-nilpotence},
respectively (notice that if an operator on a finite Hilbert space is
idempotent, one of its square roots is the operator itself). In addition, the
sum of their squares resolves the identity, $n_{c,s}^{+}+n_{c,s}^{-}=1$. Thus,
as we will see below, they can be consistently assigned the eigen-expectation-values 
$0$ or $1$, as if $\left\langle \tilde{n}_{c,s}^{\pm}\right\rangle
\mapsto\sqrt{\left\langle n_{c,s}^{\pm}\right\rangle }$.

Let us introduce the notation $\gamma_{\text{t}\sigma}\equiv\gamma_{\text{t}%
}\tilde{n}_{c}^{+}\tilde{n}_{s}^{\sigma}=\gamma_{\text{t}}\left(  \sqrt
{2}\tilde{n}_{c}^{+}\right)  \left(  \sqrt{2}\tilde{n}_{s}^{\sigma}\right)
/2$ (where the factor of $1/2$ at the end is included for a coupling-constant
rescaling in accordance with our discussion in the previous subsection). The
consistent form of the tunneling term is then more compactly rewritten as%
\begin{align}
H_{\text{tun}}  &  =-e^{ieVt}\gamma_{\text{t}\uparrow}\breve{\psi}_{l}%
\breve{\psi}_{sl}-e^{ieVt}\gamma_{\text{t}\downarrow}\breve{\psi}%
_{sl}^{\dagger}\breve{\psi}_{l}-\nonumber\\
&  \quad-e^{-ieVt}\gamma_{\text{t}\uparrow}^{\ast}\breve{\psi}_{sl}^{\dagger
}\breve{\psi}_{l}^{\dagger}-e^{-ieVt}\gamma_{\text{t}\downarrow}^{\ast}%
\breve{\psi}_{l}^{\dagger}\breve{\psi}_{sl}\,\text{.}%
\end{align}
And we can finally gauge out the applied voltage from the explicit time
dependence, as we discussed already for the conventional procedure, by using
$\psi_{l}\left(  x,t\right)  =e^{ieV\left(  t-x/v_{\text{F}}\right)  }%
\breve{\psi}_{l}\left(  x,t\right)  $. This gives%
\[
H_{\text{tun}}=-\left[  \gamma_{\text{t}\uparrow}\psi_{l}-\gamma
_{\text{t}\downarrow}^{\ast}\psi_{l}^{\dagger}\right]  \psi_{sl}-\psi
_{sl}^{\dagger}\left[  \gamma_{\text{t}\downarrow}\psi_{l}-\gamma
_{\text{t}\uparrow}^{\ast}\psi_{l}^{\dagger}\right]  \,\text{.}%
\]
Notice that we are not able to combine the fields into Majorana components, as
we did in the conventional framework, due to the spin dependence acquired by
$\gamma_{\text{t}\sigma}$. We see how this time the spin plays a role and
starts to show up clearly, as expected from our diagrammatic analysis of the problem.

\subsection{A resolution of the puzzle\label{Sec:Resolution}}

We are now ready to recompute the indirect solution to the transport problem
after debosonizing consistently. Revising the expression for the current we
find%
\begin{align*}
I  &  =-i\gamma_{\text{t}\uparrow}\left\langle \psi_{l}\left(  0,t\right)
\psi_{sl}\left(  0,t\right)  \right\rangle -i\gamma_{\text{t}\downarrow
}\left\langle \psi_{sl}^{\dagger}\left(  0,t\right)  \psi_{l}\left(
0,t\right)  \right\rangle +\\
&  \quad+i\gamma_{\text{t}\uparrow}^{\ast}\left\langle \psi_{sl}^{\dagger
}\left(  0,t\right)  \psi_{l}^{\dagger}\left(  0,t\right)  \right\rangle
+i\gamma_{\text{t}\downarrow}^{\ast}\left\langle \psi_{l}^{\dagger}\left(
0,t\right)  \psi_{sl}\left(  0,t\right)  \right\rangle \,\text{.}%
\end{align*}
We adopt the same conventions as before for the definition of the
Keldysh-Nambu spinor basis and make also the same redefinitions of the
couplings to factor out the Fermi velocity. The expression for the inverse
Green's function is like in Eq.~(\ref{Eq:invG2}) with the addition of the spin
index into the tunneling terms (which is straightforward, since all the
Nambu-off-diagonal components acquire $\sigma\!=\,\uparrow$ while the
Nambu-diagonal components go with $\sigma\!=\,\downarrow$).

It should be remarked that unpaired $\psi_{c,s}^{\left[  \dagger\right]  }$
fields do not enter in the tunneling term and appear only in the kinetic one
(as bilinears). As a result, any connected perturbative expansion in
$H_{\text{tun}}$ does not involve the \textit{charge} and \textit{spin}
sectors and the $\tilde{n}_{c,s}^{\pm}$ can be treated as c-numbers,
(restoring the Gaussianity of the problem). Due to global gauge invariance for
each lead, the final expressions involve always the squares of the $\tilde{n}%
$'s and can thus be simplified thanks to their idempotence and co-nilpotence.
An alternative equivalent calculational procedure is to set the $\tilde{n}$'s
to their different eigen-expectation-values, to do the calculation, and to
trace over all such values (not average over, because they are not exactly
conserved quantities). This second path is shorter and makes more explicit the
connection with the direct solution.

The result one gets for the I-V characteristics, by following the procedure
outlined above, is what one was hoping for:%

\begin{align}
I  &  =\frac{4\left\vert t\right\vert ^{2}}{\left(  1+\left\vert t\right\vert
^{2}\right)  ^{2}}\int_{0}^{+\infty}\left[  s_{l}\left(  \omega\right)
-\bar{s}_{l}\left(  \omega\right)  \right]  \frac{d\omega}{2\pi}\nonumber\\
&  \underset{T_{\nu}\rightarrow0}{\longrightarrow}\frac{4\left\vert
t\right\vert ^{2}eV}{\pi\left(  1+\left\vert t\right\vert ^{2}\right)  ^{2}}%
\end{align}
When comparing with the result of the direct calculation, given in
Eqs.~(\ref{Eq:IVdirectFT}) and (\ref{Eq:IVdirectZT}), the matching is now
exact and all the discrepancies are gone. Namely, (i) the $t\mapsto t/2$
correction is not required as it happened naturally courtesy of the CBCs; (ii)
the spin degeneracy arises automatically and the correct overall prefactor
arises also naturally; (iii) the extra factor of $(1+\left\vert t\right\vert
^{2})$ in the numerator is not present.

\section{Conclusion and Prospects}

\begin{figure}[ptb]
\begin{center}
\includegraphics[width=\columnwidth]{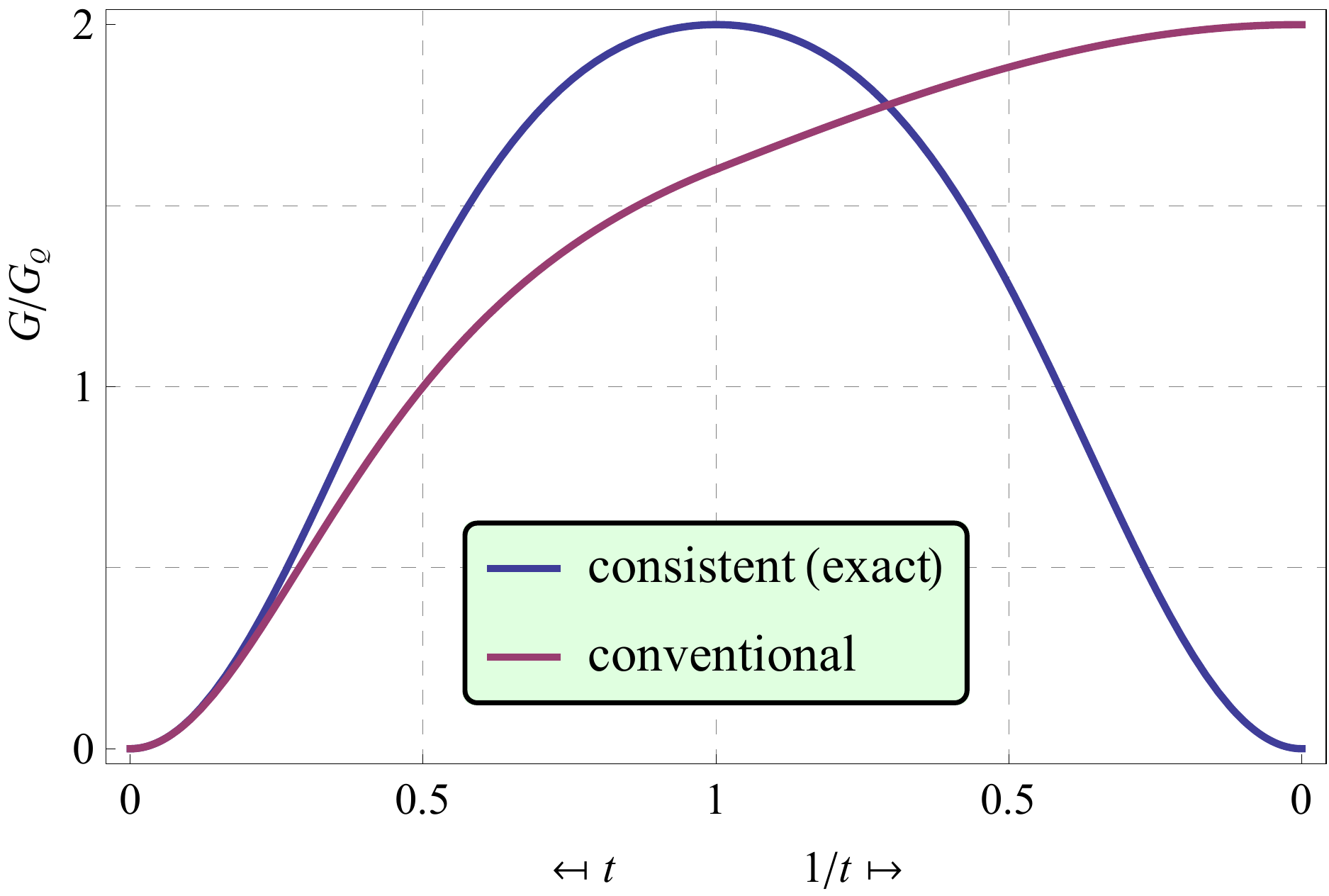}
\end{center}
\caption{Comparison of the differential conductance for the simple junction
calculated both consistently (or directly) and conventionally. Notice the
unusual convention for the horizontal axis in order to highlight the
$t\leftrightarrow1/t$ duality of the problem. The vertical axis is in units of
the single-channel quantum of conductance, $G_{Q}=e^{2}/h$, and $G=2G_{Q}$ is
the quantum limit for this problem.}%
\label{Fig_junction_dIdV}%
\end{figure}

By focusing on a case study in which bosonization is rigorously applicable
and, not less importantly, exact calculations are possible and enable detailed
comparisons, we were able to uncover some subtleties of the
bosonization-debosonization procedure that had quite strong implications.
Besides directly comparing the mathematical expressions as we have been doing,
it is instructive to compare the two results graphically. To that end, we plot
in Fig.~\ref{Fig_junction_dIdV} the two indirect solutions for the
differential conductance ($G=dI/dV$) computed conventionally and consistently.
As expected, the two results only agree in the limit of $t\rightarrow0$.
Expanding for small $t$, the two results coincide to order $\mathcal{O}\left(
t^{2}\right)  $ and start to disagree in the coefficient of the $t^{4}$ term
(with the conventional result being larger by a factor of $2$). This is as
expected from the diagrammatic analysis. Let us remark that, for $t>1$, the
differential conductance computed conventionally not only lacks the
$t\leftrightarrow1/t$ duality of the exact result, but it does not even go to
zero for $t\rightarrow\infty$. In that limit, one should have expected a
resonating-tunneling bond at the site of the junction to trap an electron (for
each spin) and thus block the passage of the current. A different way to
describe it is by appealing to a tight-binding picture. The Hamiltonian for
the two sites linked by $t$ needs to be diagonalized first when $t$ is the
largest scale in the problem. One finds bonding and anti-bonding states that,
in the $t\rightarrow\infty$ limit, will be always occupied and always empty,
respectively. The rest of the leads are relatively weakly coupled to these two
states and not able to change their fillings and thus not able to produce a
current. Instead of agreeing with this picture, the whole curve for the
conventional I-V characteristics resembles the result for a diode-like
asymmetric junction with a non-Hermitian Hamiltonian \cite{kakashvili2008a},
in which the formation of a resonating-tunneling bond is precluded by the
model. This behavior alone could have been a clear indication that there are
problems with the conventional way of calculating (even if one did not have a
direct solution to compare with).

The possible ramifications of our findings are many. A large number of the
calculations done in the past, for example any problems sharing similarities
with the one considered here (\textrm{i.e.}, involving a junction, an
impurity, or simply a boundary), will need to be reexamined critically. More
generally, problems involving backward scattering or other types of
nondiagonal interactions or processes need to be reconsidered for possible
changes. Not all past results will be significantly affected though. For
instance, on the one hand, the (weak-coupling) renormalization-group analysis of 
the effects of a \textquotedblleft classical impurity\textquotedblright\ in a
Tomonaga-Luttinger liquid \cite{kane1992,*kane1992a} requires the knowledge of
the impurity-potential beta function to leading order only, at which
consistent and conventional calculations could be expected to (at least
roughly) coincide; as our present results have shown it is the case if there
are no interactions. There will be differences, but those would be expected in
the finer details, probably appear at the next-leading order, and a
calculation would be needed to determine them. On the other hand, the
implications for the case of \textquotedblleft quantum
impurities\textquotedblright\ will be more dramatic. To put things in
perspective, the good news is that we were able to provide a clear procedure,
in the form of the $\tilde{n}$ factors, to bosonize and debosonize a large
class of models \textit{consistently}.

This paper was focused on the motivation and presentation of the formalistic
details. In the future we will look at more involved examples of greater
physical significance. We already started to reexamine some salient cases, and
in the next paper we shall focus on the important case of transport through
quantum impurities in Fermi liquids \cite{bolech2016}.

\begin{acknowledgments}
We thank P.~Kakashvili and T.~Nguyen for past interactions related to
Sec.~\ref{Sec:II} in the context of different problems. We acknowledge the
hospitality of the Kavli Institute for Theoretical Physics and the Aspen
Center for Physics (supported by the NSF under Grant Nos. PHY11-25915 
and PHYS-1066293, respectively) where part of this research was done.
\end{acknowledgments}

\appendix*

\section{Junction Thermodynamics}

The subtleties with the BdB-based mapping are quite generic and not restricted
to nonequilibrium situations. Let us briefly compute the junction
contribution to the free energy (or grand potential) and thus, indirectly, all
thermodynamic quantities. We define the junction contribution in the same way
as is done for impurity models: as the difference between the full
thermodynamic potentials with the junction closed and open, respectively.
There is no voltage applied to the junction.

\subsection{Direct Calculation}

We start with the original (old) fermions and neglect the spin which will just
give a factor of $2$ at the end. Here we follow the procedure and notations as
in Ref.~%
\onlinecite{iucci2008}%
. We will use, for convenience, a Nambu structure [otherwise we need to
introduce $\operatorname{sgn}\left(\omega_{n}\right)$ in the diagonal
entries], but we do not need to use Keldysh and, instead, we will use
Matsubara formalism and the following spinor basis:%
\[
\Psi\left(  \omega_{n}\right)  =\left(  \psi_{\text{L}}\left(  \omega
_{n}\right)  ~\psi_{\text{L}}^{\dagger}\left(  -\omega_{n}\right)
~\psi_{\text{R}}\left(  \omega_{n}\right)  ~\psi_{\text{R}}^{\dagger}\left(
-\omega_{n}\right)  \right)  ^{T}\,\text{.}%
\]
Let us use again the definition $\gamma_{\text{t}}=2v_{\text{F}}t$ and use the
standard result for the local inverse Green's function for the leads to write
the local inverse Green's function for the whole junction:%
\begin{equation}
G^{-1}\left(  \omega_{n}\right)  =-2iv_{\text{F}}%
\begin{pmatrix}
1 & 0 & it & 0\\
0 & 1 & 0 & -it\\
it^{\ast} & 0 & 1 & 0\\
0 & -it^{\ast} & 0 & 1
\end{pmatrix}
\,\text{.}%
\end{equation}
We compute the junction contribution to the thermodynamic potential via the
standard method of ``integrating over the coupling constant'':%
\begin{equation}
\Delta\Omega=\Omega-\Omega_{0}=\int_{0}^{1}\frac{d\xi}{\xi}\left\langle
\xi\,H_{\text{tun}}\right\rangle _{\xi}\,\text{.}%
\end{equation}
Introducing the action determinant%
\begin{align}
D\left(  \omega_{n},\xi\right)   &  \equiv\det G_{\xi}^{-1}\left(  \omega
_{n}\right) \label{Eq:MatActDet1}\\
&  =\left\vert t\right\vert ^{4}\xi^{4}+2\left\vert t\right\vert ^{2}\xi
^{2}+1=\left(  \left\vert t\right\vert ^{2}\xi^{2}+1\right)  ^{2}%
\,\text{,}\nonumber
\end{align}
we can use the formula%
\[
\Delta\Omega=-\int_{0}^{1}d\xi\frac{1}{\beta}\sum_{n\geqslant0}\frac
{\partial_{\xi}D\left(  \omega_{n},\xi\right)  }{D\left(  \omega_{n}%
,\xi\right)  }%
\]
Since $D\left(  \omega_{n},\xi\right)  =D\left(  \xi\right)  $ does not depend
on frequency for the problem at hand, we factor out the divergent sum and
indicate it as $\delta_{\tau=0}\equiv\frac{2}{\beta}\sum_{n\geqslant0}1$. We
have%
\begin{align}
\Delta\Omega &  =-\frac{\delta_{\tau=0}}{2}\int_{0}^{1}d\xi\,\partial_{\xi}\ln
D\left(  \xi\right) \nonumber\\
&  =-\delta_{\tau=0}\ln\left(  1+\left\vert t\right\vert ^{2}\right)
\nonumber\\
&  \underset{\times\text{ spin}}{\longrightarrow}-2\delta_{\tau=0}\ln\left(
1+\left\vert t\right\vert ^{2}\right)  \,\text{,}%
\end{align}
and we want to compare it with the result after the BdB-based transformations.

\subsection{Conventional Indirect Calculation}

Let us work in terms of the new fermions and adopt the following spinor basis:%
\[
\Psi\left(  \omega_{n}\right)  =\left(  \psi_{l}\left(  \omega_{n}\right)
~\psi_{l}^{\dagger}\left(  -\omega_{n}\right)  ~\psi_{sl}\left(  \omega
_{n}\right)  ~\psi_{sl}^{\dagger}\left(  -\omega_{n}\right)  \right)
^{T}\,\text{.}%
\]
With the same definitions, the local inverse Green's function for the junction
is%
\begin{equation}
G^{-1}\left(  \omega_{n}\right)  =-2iv_{\text{F}}%
\begin{pmatrix}
1 & 0 & -it^{\ast} & it^{\ast}\\
0 & 1 & -it & it\\
-it & -it^{\ast} & 1 & 0\\
it & it^{\ast} & 0 & 1
\end{pmatrix}
\,\text{.}%
\end{equation}
This time the action determinant reads%
\begin{equation}
D\left(  \omega_{n},\xi\right)  =4\left\vert t\right\vert ^{2}\xi
^{2}+1\,\text{,}%
\end{equation}
and applying the same formulas we find%
\begin{equation}
\Delta\Omega=-\frac{\delta_{\tau=0}}{2}\ln\left(  1+4\left\vert t\right\vert
^{2}\right)  \,\text{.}%
\end{equation}
But notice that if we \textquotedblleft correct\textquotedblright\ the
coupling constant we get%
\begin{equation}
\Delta\Omega\underset{t\mapsto t/2}{\longrightarrow}-\frac{\delta_{\tau=0}}%
{2}\ln\left(  1+\left\vert t\right\vert ^{2}\right)  \,\text{.}%
\end{equation}
We see that (i) the same \textquotedblleft correction\textquotedblright\ as in
the transport calculation is needed; (ii) we again lack an overall factor of
$4$, but (iii) the extra factor of $(1+\left\vert t\right\vert ^{2})$ is not
an issue this time (but notice that the logarithm would turn powers into factors).
Because of the last point, a perturbative analysis is not effective to
pinpoint the source of the discrepancies the way it is for transport calculations.

\textit{Nota Bene}: Working in the consistent approach and using the (non
number-eigenstates) half-filled basis $\left(  \left\vert 0\right\rangle
\pm\left\vert 1\right\rangle \right)  /\sqrt{2}$ in both the \textit{charge}
and \textit{spin} sectors, if we have $\left\langle \tilde{n}_{c}\right\rangle
=\left\langle \tilde{n}_{s}^{\sigma}\right\rangle \equiv1/\sqrt{2}$ then the
\textquotedblleft correction\textquotedblright\ of the coupling constant
reappears explicitly (but due to the use of CBCs). Moreover, tracing over the
$c$ and $s$ sectors gives the missing factor of $4$. One is thus able to
recover the direct result with a calculation which does not differ much from
the conventional one at the level of the local inverse Green's function, but
in an \textit{ad hoc} way.

\subsection{Consistent Indirect Calculation}

Let us repeat the calculation but introducing $\gamma_{\text{t}}%
\rightarrow\gamma_{\text{t}\sigma}=\gamma_{\text{t}}\tilde{n}_{c}\tilde{n}%
_{s}^{\sigma}$ (recall we divided by $2$ since we need to use CBCs). Using
the appropriately modified result for the local inverse Green's function for
the junction, one finds the following action determinant:%
\begin{align}
D\left(  \omega_{n},\xi\right)   &  =1+\left(  2t_{\uparrow}^{\ast}%
t_{\uparrow}+2t_{\downarrow}^{\ast}t_{\downarrow}\right)  \xi^{2}+\left(
t_{\downarrow}^{2}(t_{\downarrow}^{\ast})^{2}+t_{\uparrow}^{2}(t_{\uparrow
}^{\ast})^{2}\right)  \xi^{4}\nonumber\\
&  =\left(  \left\vert t\right\vert ^{2}\xi^{2}+1\right)  ^{2}\,\text{,}%
\end{align}
where the last expression is valid for either eigen-expectation-value of
$n_{s}$. One thus recovers the same expression as in the direct calculation in
the original-fermions language [cf.~Eq.~(\ref{Eq:MatActDet1})]. All the
ensuing results are thus identical. Notice the factor of $2$ for spin will be
contributed by tracing over eigenstates of $\tilde{n}_{s}$, while on the
\textit{charge} sector only the $\left\langle \tilde{n}_{c}\right\rangle =1$
subspace contributes.

\bibliography{astrings,books2015,kondo2015,hubbard2015,tunneling2015}

\end{document}